\RequirePackage{amsmath}

\documentclass[smallextended]{svjour3}   
\RequirePackage{fix-cm}

\smartqed  
\usepackage{graphicx}

\usepackage{amssymb}
\usepackage{tikz}
\usepackage{bm}
\usetikzlibrary{shapes,arrows}
\usepackage[numbers]{natbib}
\usepackage{tabularx}

\usepackage[caption=false]{subfig}	
\usepackage[labelsep=quad,skip=3pt]{caption}

\usepackage{physics}
\usepackage{multirow}

\usepackage[acronym]{glossaries}

\usepackage{pstricks}
\usepackage{pst-fill,pst-grad}
\usepackage{pst-text}
\usepackage{pst-uml}
\usepackage{pst-3d,pst-coil,pst-eps,pst-node}

\renewcommand\paragraph{\@startsection{paragraph}{4}{\z@}%
            {-2.5ex\@plus -1ex \@minus -.25ex}%
            {1.25ex \@plus .25ex}%
            {\normalfont\normalsize\bfseries}}
\usepackage{xcolor}
\usepackage[framemethod=TikZ]{mdframed}
\definecolor{ikmgray}{HTML}{5E5E5E}
\definecolor{ikmgreen}{HTML}{C9DA2B}
\newmdenv[frametitle={},
middlelinecolor=ikmgreen,
middlelinewidth=0pt,
backgroundcolor=ikmgray!20,
roundcorner=2pt,
bottomline=false,
leftline=true,
topline=false,
rightline=false,
skipabove=10pt,
skipbelow=10pt,
leftmargin=10pt,
rightmargin=10pt,
innerleftmargin=10pt,
innerrightmargin=10pt,
innertopmargin=10pt,
innerbottommargin=10pt]{Algorithmus}
\usepackage[labelfont=bf, 
format=plain, 
labelsep=endash, 
justification=raggedright 
]{caption}
\DeclareCaptionType{kasten}[Box]
\newacronym{gcd}{GCD}{Greatest Common Divisor} 
\newacronym{lcm}{LCM}{Least Common multiple}
\newacronym{olhd}{OLHD}{Optimal Latin Hypercube Design}
\newacronym{svm}{SVM}{Support Vector Machines}
\newacronym{pr}{PR}{Polynomial Regression}
\newacronym{rbfn}{RBFN}{Radial Basis Function Network}
\newacronym{nn}{NN}{Neural Networks}
\newacronym{mae}{MAE}{Mean Absolute Error}
\newacronym{lf}{LF}{Low-Fidelity}
\newacronym{hf}{HF}{High-Fidelity}
\newacronym{doe}{DOE}{Design of experiments}
\newacronym{ok}{OK}{Ordinary Kriging}
\newacronym{uk}{UK}{Universal Kriging}
\newacronym{hk}{HK}{Hierarchical Kriging}
\newacronym{pls}{PLS}{Partial Least Squares}
\newacronym{plsok}{PLSOK}{Partial Least Squares Ordinary Kriging}
\newacronym{plshk}{PLSHK}{Partial Least Squares Hierarchical Kriging}
\newacronym{mipt}{MIPT}{MC-intersite-proj-th }
\newacronym{msd}{MSD}{Maximin Scaled Distance}
\newacronym{cv}{CV}{Cross-Validation}
\newacronym{cvv}{CVV}{Cross-Validation Variance}
\newacronym{cvvor}{CVVOR}{Cross-Validation-Voronoi}
\newacronym{cdm}{CDM}{Crowding Distance Metric}
\newacronym{ssa}{SSA}{Smart Sampling Algorithm}
\newacronym{gcv}{GCV}{Generalized Cross-Validation}
\newacronym{mse}{MSE}{Mean-Squared Error}
\newacronym{mmse}{MMSE}{Maximum Mean-Squared Error}
\newacronym{imse}{IMSE}{Integrated Mean-Squared Error}
\newacronym{ame}{AME}{Adaptive Maximum Entropy}
\newacronym{ei}{EI}{Expected Improvement}
\newacronym{cdf}{CDF}{Cumulative Distribution Function}
\newacronym{pdf}{PDF}{Probability Distribution Function}
\newacronym{wei}{WEI}{Weighted Expected Improvement}
\newacronym{awei}{AWEI}{Adaptive Weighted Expected Improvement}
\newacronym{eigf}{EIGF}{Expected Improvement for Global Fit}
\newacronym{haed}{HAED}{Hierarchical Adaptive Experimental Design}
\newacronym{gek}{GEK}{Gradient-Enhanced Kriging}
\newacronym{nurbs}{NURBS}{Non-Uniform Rational B-Spline}
\newacronym{lola}{LOLA}{Local Linear Approximation}
\newacronym{qbc}{QBC}{Query-By-Committee}
\newacronym{masa}{MASA}{Mixed Adaptive Sampling Algorithm}
\newacronym{lhd}{LHD}{Latin Hypercube Design}
\newacronym{tplhd}{TPLHD}{Translational Propagation Latin Hypercube Design}
\newacronym{rmse}{RMSE}{Root Mean-Squared Error}
\newacronym{rmae}{RMAE}{Relative Maximum Absolute Error}
\newacronym{cvd}{CVD}{Cross-Validation Distance}
\newacronym{le}{LE}{Lyapunov Exponents}
\newacronym{lle}{LLE}{Largest Lyapunov Exponent}
\newacronym{epe}{EPE}{Expected Prediction Error}
\newacronym{mepe}{MEPE}{Maximizing Expected Prediction Error}
\newacronym{loocv}{LOCVV}{leave-one-out cross-validation}
\newacronym{gmse}{GMSE}{Generalized Mean Square Cross-Validation Error}
\newacronym{sfcvt}{SFCVT}{Space-Filling Cross Validation Tradeoff}
\newacronym{ace}{ACE}{ACcumulative Error}
\newacronym{doi}{DOI}{Degree-of-Influence}
\newacronym{blup}{BLUP}{Best linear unbiased predictor}
\newacronym{mle}{MLE}{Maximum likelihood estimation}
\newacronym{de}{DE}{Differential evolution}
\newacronym{mivor}{MiVor}{Monte Carlo-Intersite Voronoi}
\newacronym{qoi}{QoI}{Quantity of Interest}
\newacronym{mob}{MoB}{Mass-on-Belt}
\begin{document}

\title{Surrogate model approach for investigating the stability of a friction-induced oscillator of Duffing's type
}

\titlerunning{Kriging approach to investigate stick-slip instability}        

\author{Jan N. Fuhg         \and
        Am\'{e}lie Fau  
}


\institute{Jan N. Fuhg (Corresponding author) \at
Institute of continuum mechanics\\
              Leibniz Universit{\"a}t Hannover \\
              Appelstra{\ss}e 11 \\
              30167 Hannover, Germany\\
              \email{fuhg@ikm.uni-hannover.de} 
             \\
Tel.:  +49 (0)511.7 62 - 22 85 \\
Fax: +49 (0)511.7 62 - 54 96
                \\
\\
Am\'{e}lie Fau \at
Institute of mechanics and computational mechanics \\
              Leibniz Universit{\"a}t Hannover \\
              Appelstra{\ss}e 9A \\
              30167 Hannover, Germany\\
}

\date{Received: date / Accepted: date}

\maketitle

\begin{abstract}

Parametric studies for dynamic systems are of high interest to detect instability domains. This prediction can be demanding as it requires a refined exploration of the parametric space due to the disrupted mechanical behavior. In this paper, an efficient surrogate strategy is proposed to investigate the behavior of an oscillator of Duffing's type in combination with an elasto-plastic friction force model. Relevant quantities of interest are discussed. Sticking time is considered using a machine learning technique based on Gaussian processes called kriging. The largest Lyapunov exponent is proposed as an efficient indicator of non-regular behavior. This indicator is estimated using a perturbation method. A dedicated adaptive kriging strategy for classification called MiVor is utilized and appears to be highly proficient in order to detect instabilities over the parametric space and can furthermore be used for complex response surfaces in multi-dimensional parametric domains. 
 
\textcolor{blue}{}
\end{abstract}



\section{Introduction}
\label{intro}

Stick-slip instability can be observed when two rigid bodies in contact slide on each other at low relative velocity. Then, intermittent vibratory effects may occur in form of variations of the frictional force and relative sliding velocity, with a characteristic sawtooth time-displacement curve, corresponding to stick and slip phases \cite{rabinowicz1956stick}. These dynamic instabilities, also called stick-slip motion result in self-excited vibrations and drastic decreases in performance of some machine parts. Oscillating systems excited by dry friction are frequently encountered in many practical applications, including for instance brake systems \citep{barton2004braking,ashraf2017investigation}, hydraulic cylinders \citep{owen2003reduction}, gears \cite{wu2017comparisons} or bearing \cite{rubio2007structural}. Therefore the extensive analysis of these non-smooth dynamic systems to detect instabilities is a crucial point in engineering design.

Linear spring mass systems placed on a moving belt, commonly referred to as \gls{mob} systems, have been generally utilized as simple mathematical models to investigate stick-slip system, see e.g. \cite{stelter1990stick,hinrichs1997dynamics,galvanetto1999dynamics,jimenez2007two}. To provide better accuracy, nonlinear mathematical representations have recently been considered \citep{devarajan2017analytical}. For instance, the bifurcation behavior of the non-linear \gls{mob} system has been investigated in \cite{santhosh2016discontinuity,awrejcewicz2005stick}. A Duffing's type oscillator has been studied  analytically in \cite{devarajan2017analytical} in order to obtain expressions for stick-slip and pure-slip vibration amplitudes and frequencies. An estimation method for the spectrum of \gls{le}s proposed by \citep{balcerzak2018spectrum} is employed in \cite{pikunov2019numerical} to analyze the stability of a discontinuous \gls{mob} system.

To avoid computing and analyzing the behavior in the time domain, some indicators of instabilities have been proposed in the literature, such as Kolmogorov entropy \cite{benettin1976kolmogorov}, correlation dimension \cite{grassberger1983characterization} and sticking time \cite{lima2015stick}. \gls{le}s have been stated by some authors as the most significant indicator for chaotic motion such as in \citep{kocarev2006discrete}. They quantify the rate of separation of infinitesimally close trajectories and the sensitive dependence on initial conditions exhibited by the system. In practical applications the determination of the \gls{lle} is sufficient. If the number is non-positive, then the system is stable. On the contrary, a positive \gls{lle} indicates a system with chaotic behavior.  Unfortunately, analytic determination of \gls{le}s is possible only for simple linear dynamic systems, see \cite{balcerzak2018spectrum}, and numerical determination of their value is a challenge. In recent years several numerical strategies have been suggested for estimating LEs in case of non-smooth dynamic systems. A thorough review can be found in \cite{awrejcewicz2018quantifying}. Recently an approximate numerical method based on the estimation of the Jacobian matrix by perturbation method of the initial orthogonal vectors involving an Euler forward scheme has been proposed \cite{balcerzak2018spectrum}.

In addition, engineers want to prevent instabilities for a large scope of parameters, or scenarios which may be encountered during the working life of the mechanical systems. However the numerical analysis of the \gls{lle} in complex systems is computationally expensive, limiting the design space exploration and the optimization of a system at hand. Therefore statistical approximations or surrogate models emulating the \gls{lle} over a parametric input space appear very interesting to estimate the response on vast parametric space at low cost.
Surrogate model approaches such as polynomial response, see e.g. \cite{kleijnen2017regression}, kriging also known as Gaussian process regression \citep{kleijnen2009kriging,williams1996gaussian}, support vector regression \citep{clarke2005analysis} or radial basis function models \citep{park1991universal} have been developed and continuously improved in recent years.
Kriging is an accurate interpolative Bayesian surrogate modeling technique, which is stated as the most intensively investigated surrogate approach \cite{jiang2017adaptive}. Originally proposed by the geostatistician Krige \cite{krige1951statistical}, it has been utilized for deterministic \citep{sacks1989design} and random simulations models \citep{van2003kriging}. The idea of a kriging model is to provide an estimation of the response of the system on the whole parametric domain from the restricted available information provided by some observations. Two perspectives can be explored. One is generating an accurate surrogate model from the information given \textit{a priori}. The second is to inform about the localization of crucial areas in the parametric space for the problem of interest, i.e. subdomains in which more information should be gained, if further exploration is possible. In light of recent contributions proposed for dynamical analysis and machine learning tools, the goal of this contribution is to propose surrogate strategies to investigate the oscillatory behavior and efficiently detect the risk of instabilities.

Thus, in this paper surrogate model construction to identify the parametric subdomains where dynamic instabilities may occur is investigated. Pertinent instability indicators are discussed, different metamodeling techniques are contemplated. Among them, kriging appears the most powerful for the problem of interest. Numerical challenges and capacities offered by kriging for the estimation of the sticking time and \gls{lle} analysis in a Duffing's type oscillator with an elasto-plastic friction force model are exposed. More particularly, various adaptive schemes are evaluated in this context.

The paper is organized as follows. In section \ref{section:dyn_prob} a dynamic problem non-linear contact behavior is introduced, as well as the accompanying challenges due to the possible unstable behavior. Two possible indicators proposed by the literature to proficiently investigate the system behavior are exposed in section \ref{sec:inst_indic}. A brief introduction to LE as well as most recent works which allow to estimate the \gls{lle} for discontinuous systems in a robust numerical framework is also presented. In section \ref{sec:surrogate}, an overview of the kriging approach is given. Section \ref{sec:Kriging_dyn} represents the core of this contribution, with a detailed review of the ability and challenge of the surrogate model approach to investigate unstable dynamic behaviors.

\section{Dynamic problem of interest}
\label{section:dyn_prob}

\begin{figure}[htp!]
\centering
  \includegraphics[scale=0.35]{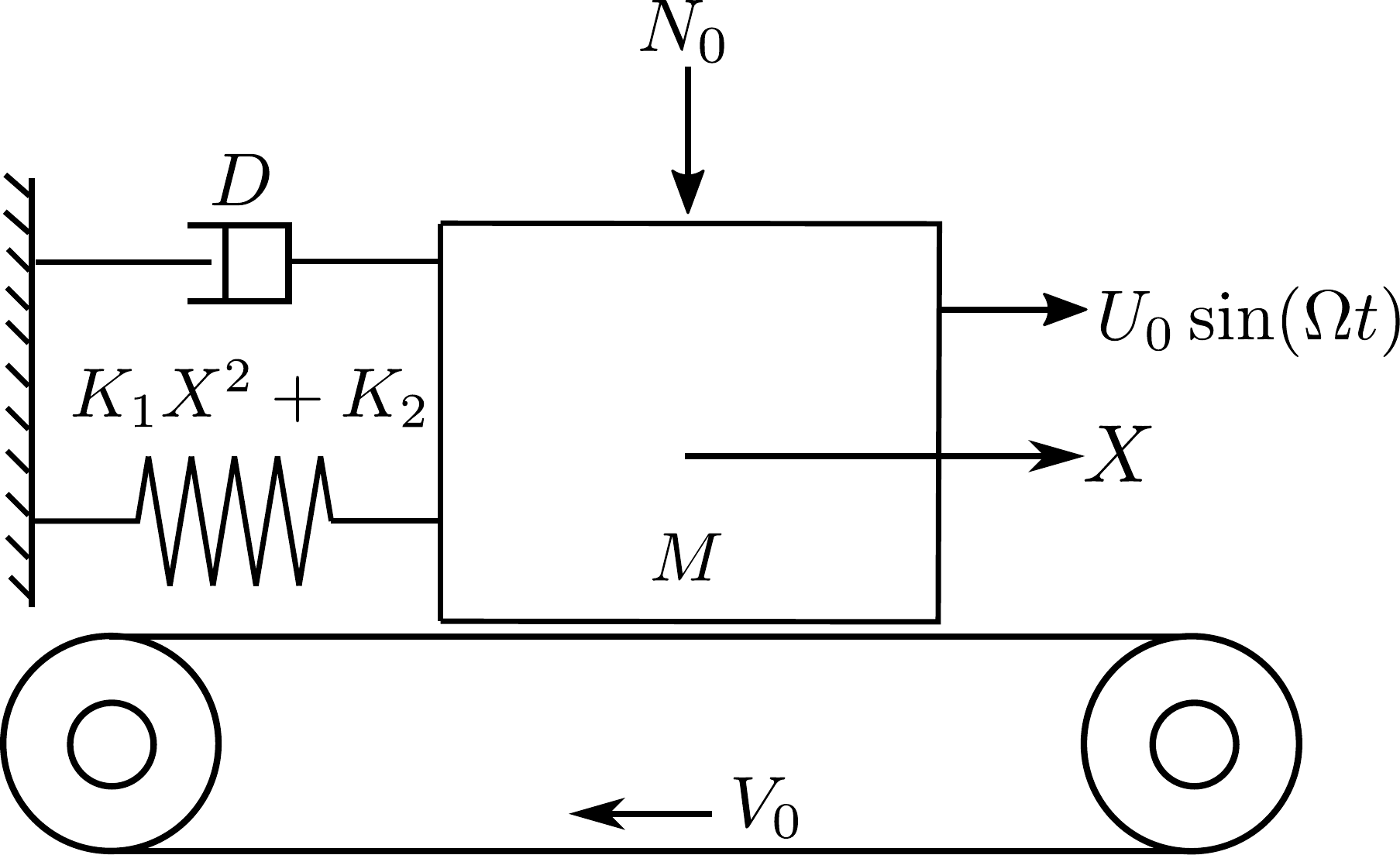}
\vspace{0.5cm}
\caption[Scheme of the analyzed nonlinear mass-on-belt system]{Scheme of the analyzed nonlinear mass-on-belt system.}
\label{fig:Application}       
\end{figure}

The mechanical reference problem is a Duffing's type oscillator with a damping term, as illustrated in Figure \ref{fig:Application}. A rigid body of mass $M $ is placed on a moving belt with a constant velocity $V_{0}$. The displacement of the body over time $t$ is denoted by $X(t)$. The body's movement is restricted by a  nonlinear spring of stiffness $K_{1} X^{2} + K_{2}$ and a dashpot with damping coefficient $D$ parallel to
the spring.
The relative velocity between the belt and the body is denoted by $V_{R}(t)$, which equals to $ V_{0} - \dot{X}(t) $ with $\dot{X}(t)$ representing the first derivative of the body's displacement with respect to time, i.e. its velocity. 
A time-dependent harmonic force $U(t) = U_{0} \sin (\Omega t)$ with amplitude $U_{0} $ and angular frequency $\Omega $ as well as a constant normal load $N_{0}$ is applied to the mass. 

\begin{figure}[htp!]
\centering
\includegraphics[scale=0.4]{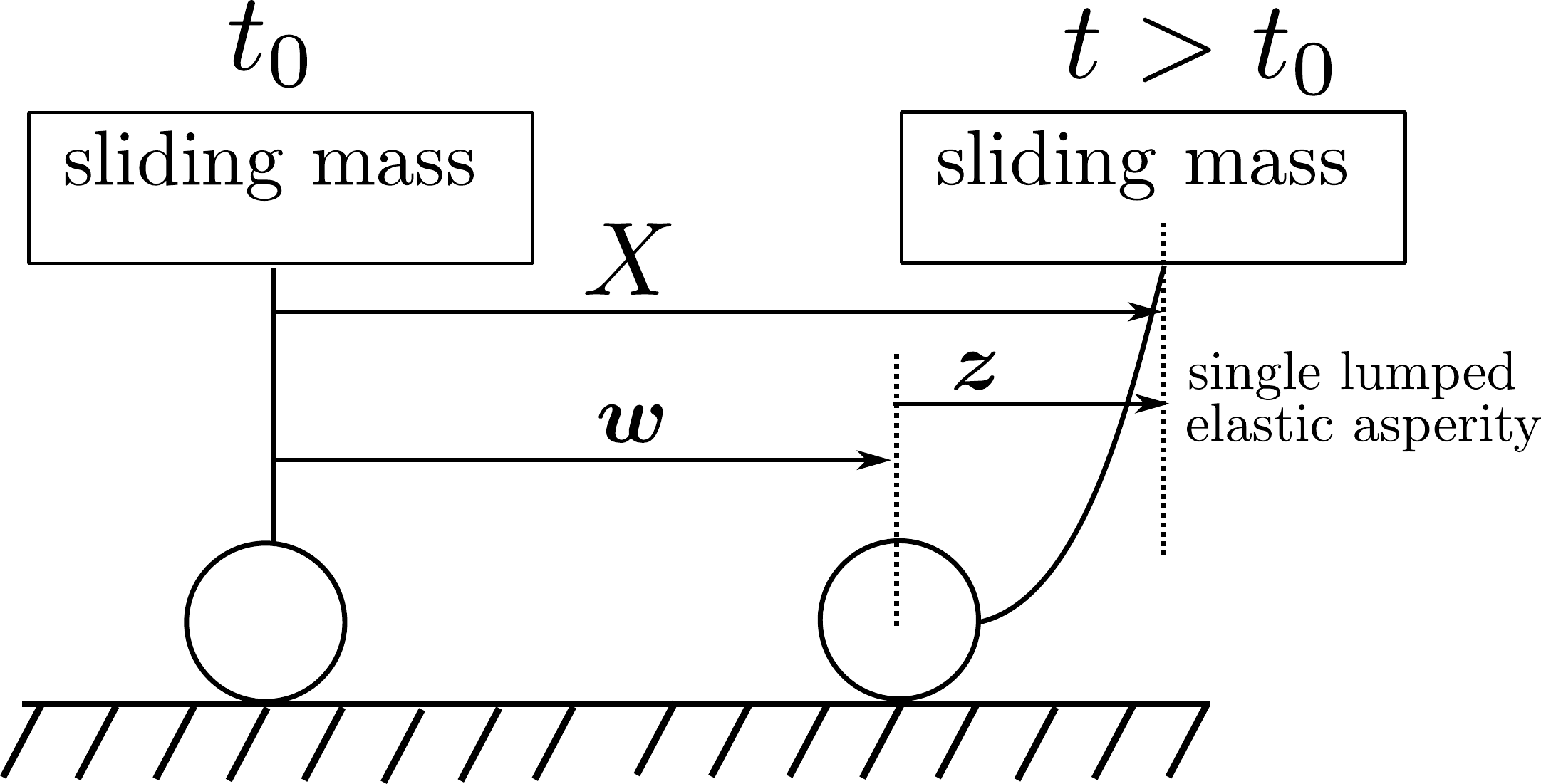}
\vspace{0.5cm}
\caption[Physical analogy of the elasto-plastic model]{Physical analogy of the elasto-plastic model representing the contact behavior - the block displacement $X$ over the surface is broken down into an elastic component $z$ and a plastic part $w$, modified from \cite{dupont2000elasto} }\label{fig::elasto-plastic}
\end{figure}

The equation of motion of the system reads 
\begin{equation}
\begin{aligned}
M \ddot{X}(t) = - D \dot{X}(t)  - K_{1} X^{3}(t) -K_{2} X(t)   + F_{R}(V_{R}) + U_{0} \sin(\Omega t) \, \text{,}
\end{aligned}
\end{equation}
where $\ddot{X}(t)$ is the second derivative of the body's displacement with respect to time, i.e. the body acceleration and $F_{R}(V_{R}) = N_{0} f_{R}(V_{R})$ is the friction force. $f_{R}(V_{R})$ is called the friction function which denotes the friction force per unit of normal load; it is a function chosen by the user. An overview of the common choices proposed in the literature is given in \cite{pennestri2016review}. Generally, friction force models are classified either as static or dynamic \citep{olsson1998friction}, where in a dynamic model the friction force does not only depend on the relative velocity between the bodies but also on other state variables. A review of different friction force models for dynamic analysis has been presented in \cite{marques2016survey}.
Here, a dynamic model proposed in \cite{dupont2002single} is employed. It belongs to the family of elasto-plastic models, which were initially introduced by 
\cite{prandtl1924spannungsverteilung} to model the constitutive behavior of deformable bodies. Internal variables are used to represent the non-linear contact behavior. The displacement $X$ of the body is assumed to be the sum of an elastic, i.e. memoryless, contribution denoted by $z$ and a plastic, and so history-dependent, part $w$ such that
\begin{equation}
X = z + w \, \text{.}
\end{equation} 
A physical analogy of this model is depicted in Figure \ref{fig::elasto-plastic}. During the sticking phase, the plastic displacement is constant, whereas while slipping the elastic component is assumed to be fixed. 

\begin{figure}[htp!]
\centering
\vspace{0.2cm}
\includegraphics[scale=0.4]{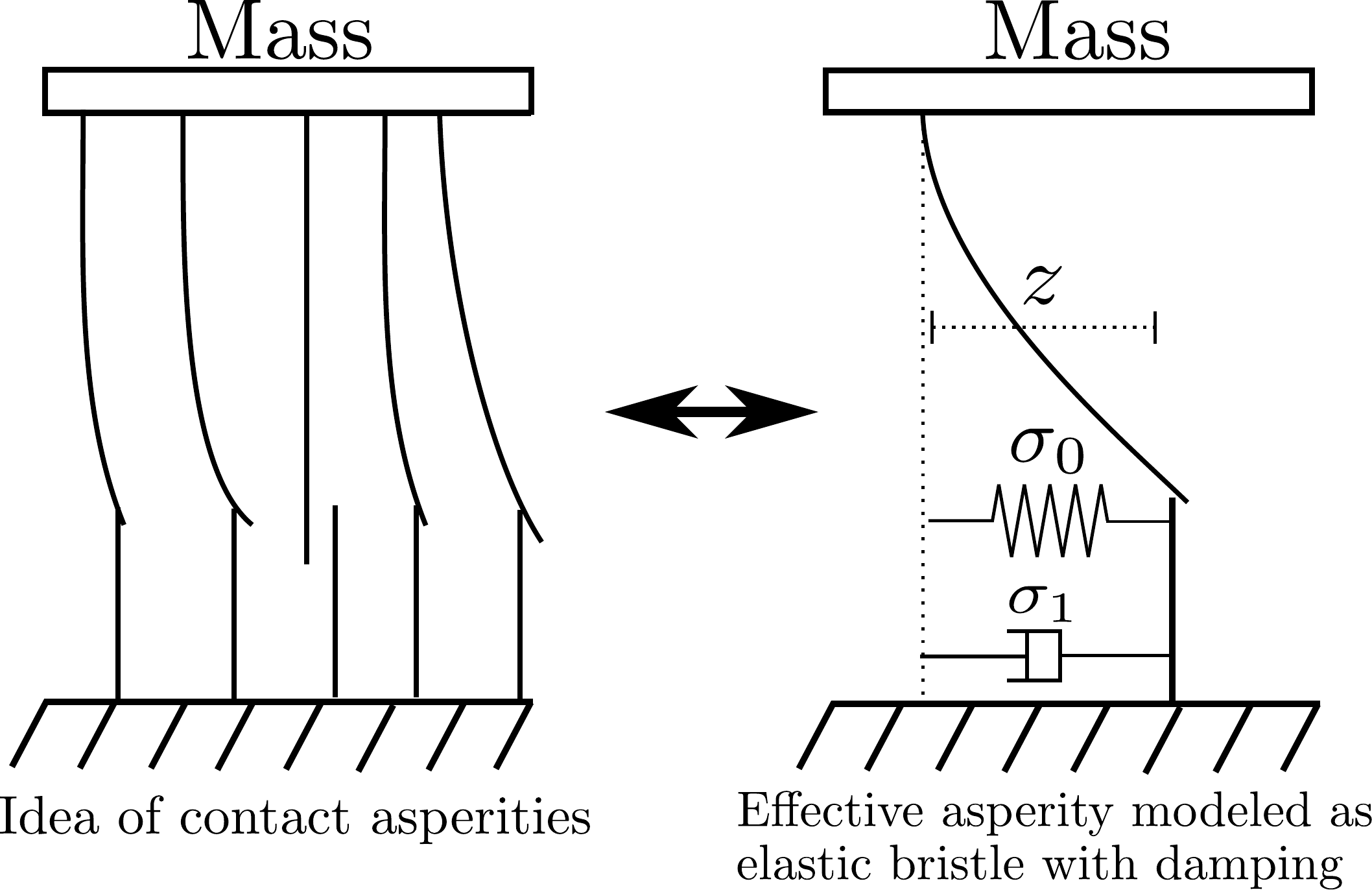}
\vspace{0.5cm}
\caption[Basic concept of friction force formulation]{Basic concept of friction force formulation. Asperities in the contact surface are modeled as elastic bristles with damping. For simplicity the LuGre model considers an average bristle deflection $z$ for the determination of the friction force.}\label{fig::bristle}
\end{figure}

The friction model, as illustrated in Figure \ref{fig::bristle}, is based on the idea that any deformation of the asperities is associated with a corresponding friction force $F_{R}$. By analogy with plasticity, the history of the model is characterized by the internal variable $z$, which represents the bristle deflection. 
The friction function is defined as
\begin{equation}
\begin{aligned}
f_{R}(V_{R},z)= \sigma_{0} z + \sigma_{1} \dot{z} + \sigma_{2} V_{R},
\end{aligned}
\end{equation}
which can be compared to the force model of the classical LuGre model \cite{de1995new}. The variable $\sigma_{0}$ denotes the bristle stiffness, $\sigma_{1}$ is the average bristle damping coefficient and $\sigma_{2}$ is a viscous component of the friction force. 

Consider the following definition
\begin{equation}
\text{sgn} \left( V_{R} \right) =  \begin{cases}
\frac{V_{R}}{\norm{V_{R}}}, & \text{if} \, V_{R} \neq 0, \\
0, & \text{if} \,  V_{R} = 0,
\end{cases}
\end{equation}
the velocity of bristle deflection $\dot{z}$ has been defined by  \cite{dupont2000elasto} as
\begin{equation}
\begin{aligned}
\dot{z} = \left(1 - \alpha(z, V_{R}) \dfrac{\sigma_{0}}{g(V_{R})} z \cdot \text{sgn} \left( V_{R} \right) \right) V_{R} \, \text{,}
\end{aligned}
\end{equation}
where $\alpha(z, V_{R})$ is a function, which is utilized to capture stiction effect. It depends on the relative velocity see e.g. \cite{townsend1987effect}, since it defines the elastic deformation until the breakaway force of the system is exceeded, i.e.
\begin{equation}
\begin{aligned}
\alpha(z, V_{R}) = \begin{cases}
\alpha(z) , & \text{if} \;  V_{R} \cdot z \geq 0 \, \text{,} \\
0 , & \text{if} \:  V_{R} \cdot z < 0 \, \text{,}
\end{cases}
\end{aligned}
\end{equation}
with
\begin{equation}
\begin{aligned}
\alpha(z)= \begin{cases}
0 , & \text{if} \, 	 z \leq z_{ba}, \\
\frac{1}{2} \left( \sin \left( \pi \dfrac{\norm{z}  - \frac{z_{max} + z_{ba}}{2} }{z_{max} - z_{ba}} \right) + 1 \right) , & \text{if} \,  z_{ba} < \norm{ z} < z_{max}, \\
1 , & \text{if} \,  z_{max} \leq \norm{ z } \, \text{.}
\end{cases}
\end{aligned}
\end{equation}
The maximum bristle deflection is denoted by $z_{max}$, and the bristle deflection for breakaway condition as $z_{ba}$.
The function $g(V_{R})$ models the Stribeck effect, which refers to an increase of the friction coefficient when the relative contact velocity increases \cite{stribeck1902wesentlichen}.
It is given by
\begin{equation}
g(V_{R}) = N_{0} \left( \mu_{k} + (\mu_{s}-\mu_{k}) \exp \left( - \frac{V_{R}^{2}}{V_{S}^{2}} \right)   \right),
\end{equation}
where $V_{S}$ is a parameter called the characteristic Stribeck velocity, and $\mu_{s}$ and  $\mu_{k}$ denote the static and kinetic friction coefficients, respectively.
Therefore, the complete equation of motion of the system in state-space form with the modified elasto-plastic friction model reads 
\begin{equation}\label{eq::dynamic_system}
\begin{cases}
\frac{dX}{dt} &= \dot{X}, \\
\frac{d^{2} X}{d t^{2}} &= - \frac{D}{M} \dot{X}  - \frac{K_{1}}{M} X^{3} - \frac{K_{2}}{M} X   +  \frac{N_{0}}{M} f_{R}(V_{R}) + \frac{U_{0}}{M} \sin(\Omega t), \\
\frac{d z}{d t} &= \left(1 - \alpha(z, V_{R}) \dfrac{\sigma_{0}}{g(V_{R})} z \cdot \text{sgn} \left( V_{R} \right) \right) V_{R} \, \text{,}
\end{cases}
\end{equation}
which can be solved, e.g. via numerical integration. 

\begin{figure}[htp]
\centering
   \subfloat[Time response]{\label{fig::Time}
\includegraphics[scale=0.3]{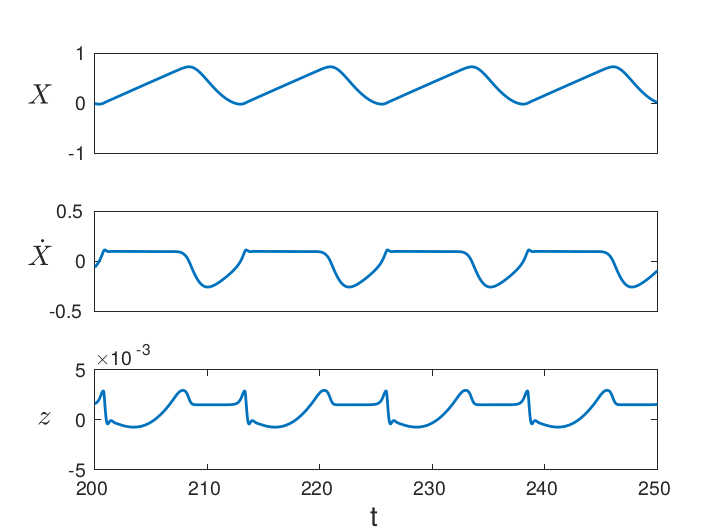}}
\subfloat[Phase portrait]{\label{fig:PhaseResponse}
 \includegraphics[scale=0.3]{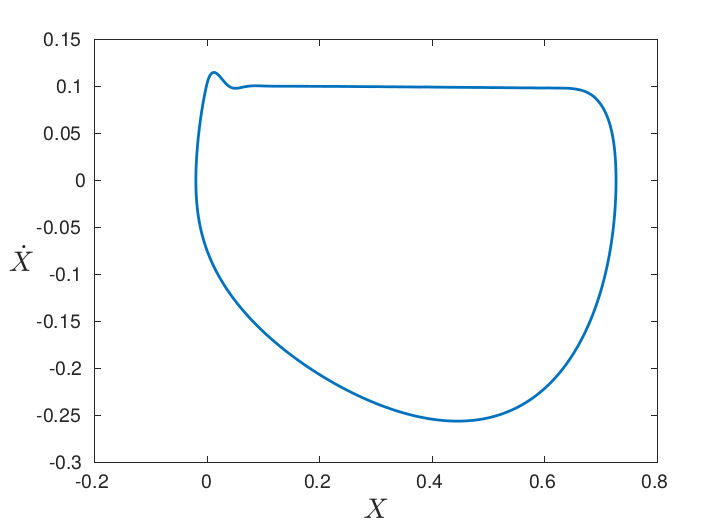}}
   \caption[Time response and phase-space for application]{Time response and phase portrait for the nonlinear mass-on-belt system with $X$ in $\text{m}$, $\dot{X}$ in $\text{m}.\text{s}^{-1}$, $t$ in $\text{s}$, $\Omega = 0.5 \, \text{rad}.\text{s}^{-1}$,
$M = 1 \, \text{kg}$,
$V_{0} = 0.1 \, \text{m}.\text{s}^{-1}$,
$D = 0.0 \, \text{Ns}.\text{m}^{-1}$, 
$K_{1} = 1\, \text{N}.\text{m}^{-3}$,
$K_{2} = 0.0 \, \text{N}.\text{m}^{-1}$, 
$\mu_{s} = 0.3$,
$\mu_{k} = 0.15$,
$V_{s} = 0.1 \, \text{m}.\text{s}^{-1}$,
$U_{0} = 0.1 \,\text{N}$,
$N_{0} = 1.0 \, \text{N}$,
$\sigma_{0} = 100.0 \, \text{N}.\text{m}^{-1} $,
$\sigma_{1} = 10.0 \,  \text{Ns}.\text{m}^{-1}$ and
$\sigma_{2} = 0.1 \, \text{Ns}.\text{m}^{-1}$.}\label{fig:TimeAndPhase}
\end{figure}

From the equations of motion, the dynamic behavior of the system can be analyzed. For instance, the time-response behavior plotted in Figure \ref{fig::Time} shows the cyclic unstable behavior in time of the mass via the typical cyclic evolution of its displacement and its velocity. The cyclic unstable behavior can also be analyzed via the phase portrait as shown in Figure \ref{fig:PhaseResponse}. Analysis of the stick-slip motion using phase space is detailed in \cite{galvanetto1999dynamics}.

\noindent The temporal analysis has been historically established. However, to automatically scan the parametric space and benefit from modern machine learning tools, it appears interesting to analyze the dynamic behavior via single scalar values, used as quantities of interest for the surrogate model.

\section{Quantities of interest to analyze the dynamic system} 
\label{sec:inst_indic}
Among the different scalar values proposed in the literature to investigate the dynamic system two are considered, namely the sticking time and the largest Lyapunov exponent.

\subsection{Sticking time}
The dynamic cyclic behavior as represented in Figure \ref{fig::Time} can be decomposed into sticking and slipping modes. In \cite{lima2015stick} it has been proposed to investigate the oscillation behavior of dry-friction oscillator using the percent of time in which the body remains in stick mode as \gls{qoi}. They suggest to improve properties e.g. endurance of mechanical systems by understanding the role of acting parameters on that \gls{qoi}. 

In the following, the sticking time of the \gls{mob} is defined as the time during which the relative velocity $V_{R}(t) = V_{0} - \dot{X}(t)$ remains below the threshold of $10^{-4}\,m.s^{-1}$. It is numerically estimated by integrating equation (\ref{eq::dynamic_system}) with a six-stage, fifth-order, Runge-Kutta method with variable step-size. In order to avoid localized transient behavior, sticking time is analyzed on a time window between $150$ and $250$ seconds.
\begin{figure}
    \centering
    \includegraphics[scale=0.5]{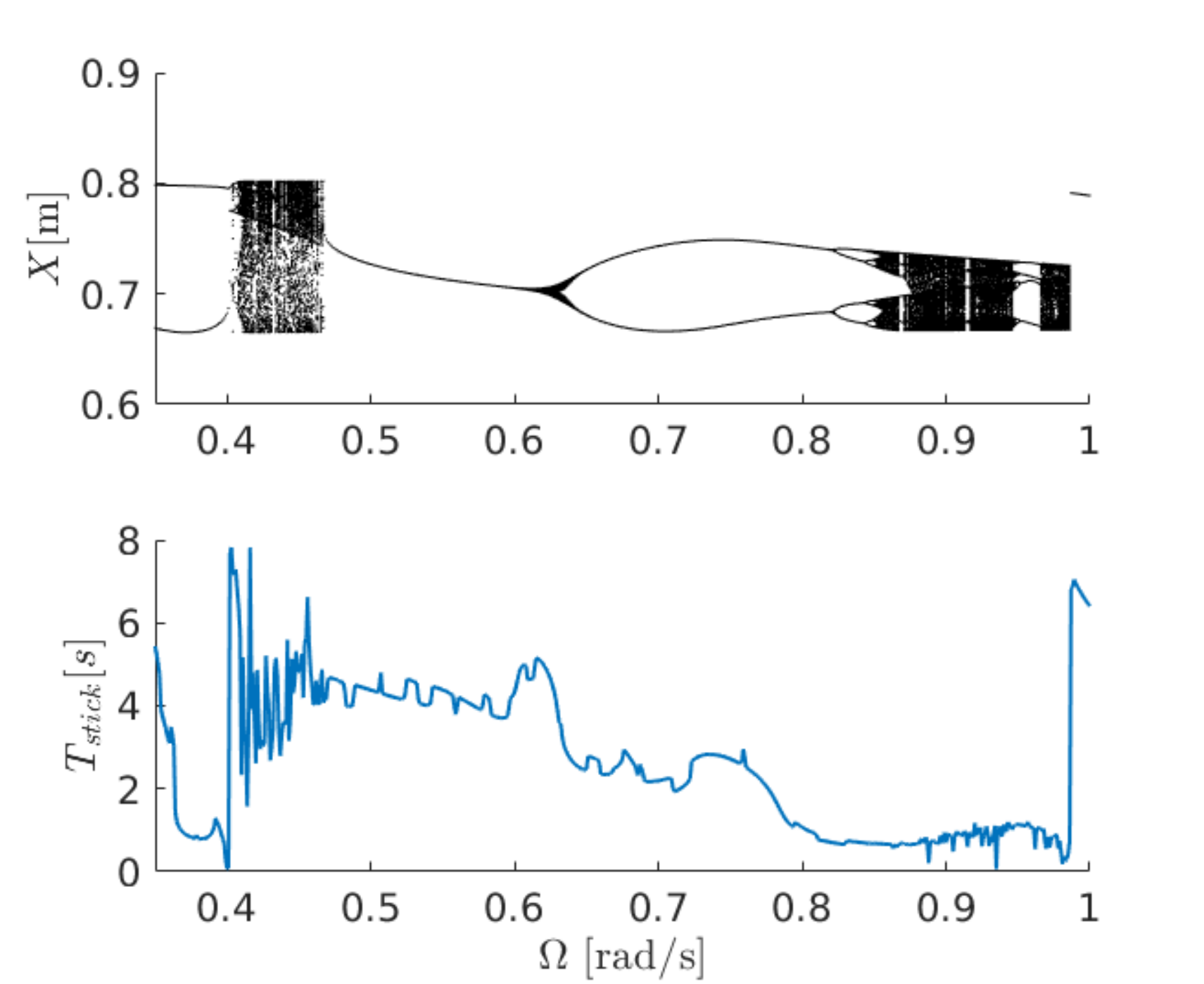}
    \caption{Sticking time as indicator for chaotic motion. Upper image: Bifurcation diagram over angular frequency. Lower image: Sticking time over angular frequency.}
    \label{fig:stick_time_as_qoi}
\end{figure}
The analysis of this \gls{qoi} is, however, not evident because as shown in Figure \ref{fig:stick_time_as_qoi}, the value of the sticking time cannot be directly associated with either stable or unstable behavior. Lima and Sampaio  pointed out that both the duration of the sticking mode and the cumulative sticking time lack attention in the literature \cite{lima2015stick}. However, this analysis is not the scope of this contribution, here the sticking time is simply considered as a candidate \gls{qoi} to be represented by the metamodel.


\subsection{Largest Lyapunov exponent}

The second \gls{qoi} aims at providing an instability indicator to break down the parametric space into stable and unstable domains.

Consider a general equation of motion of an $N$-dimensional time-continuous system recast into first order differential equation system of the form
\begin{equation}\label{eq::n_dim_system}
\dot{\bm{x}} = \bm{f} (\bm{x}) \, \text{,}
\end{equation}
with the state-space vector $\bm{x} \in \mathbb{R}^{N}$ and $\bm{f}$ being a set of $N$ functions i.e. $\bm{f}= \left[f_{1}(\bm{x}), \ldots f_{N}(\bm{x}) \right]$. Consider an $N$-dimensional sphere of initial conditions. With time the sphere evolves into an ellipsoid whose principal axes contract or expand with rates governed by the spectrum of \gls{le}s denoted by $\left\lbrace \lambda_{i} \right\rbrace_{i \in \left[ 1,N\right] }$. \gls{le}s introduced by \cite{oseledec1968multiplicative} are defined as the average divergence or convergence rates of nearby orbits in state space. A positive exponent indicates local instability in this particular direction and hence characterizes chaotic motion, see \cite{rosenstein1993practical}.  The presence of a positive \gls{lle} indicates chaos, whereas negative values are characteristic of regular motion. Therefore to ensure the stability of the system, computing only the \gls{lle} is sufficient. The LLE is considered as one of the most useful tools to characterize the stability of a dynamic system in \cite{kocarev2006discrete}. Therefore, it should turn out as a proficient \gls{qoi} for building an univariate informative surrogate model.  

However, analytic calculation of the \gls{lle} is mostly restricted to simple linear systems. For complex systems robust numerical techniques can be employed to estimate the LLE, see e.g. \cite{shimada1979numerical},\cite{benettin1980lyapunov}, or \cite{kantz1994robust}. 

\subsubsection{Numerical estimation of the Lyapunov exponents}\label{sec::EstimationLLE}

When the investigated dynamic system is smooth with an analytically obtainable Jacobian matrix, the \gls{lle} can be calculated with e.g. the algorithm given in \cite{wolf1986quantifying}, in which the system of equations is solved for $N$ nearby initial conditions and correct state-space orientation is maintained by repeatedly orthonormalizing the corresponding set of vectors using the Gram-Schmidt procedure. However, the dry friction problem at hand is a non-smooth system. The Jacobian matrix cannot be determined or is strongly ill-conditioned, which leads to significant numerical problems when employing the algorithm proposed by \cite{wolf1986quantifying}. 
To overcome this obstacle, a novel method has recently been presented in \cite{balcerzak2018spectrum}, which estimates the Jacobian matrix by a truncated Taylor series expansion with small orthogonal perturbations. The spectrum of Lyapunov exponents has been estimated from that numerical method in \cite{pikunov2019numerical} to analyze the stability of a nonlinear mass-on-belt system. It is shortly summarized as follows:

Consider a discretization in time given by the set $\lbrace t_{i} | i=1, \ldots, n \rbrace$. Let $x(t_{i})$ be denoted by $x_{i}$.
The discrete dynamical system of the mapping $\bm{f}$ introduced in equation (\ref{eq::n_dim_system})
at time $t_{i}$ reads
\begin{equation}\label{eq::gen_form}
\bm{x}_{i+1} = \bm{G}(\bm{x}_{i}),
\end{equation}
where $\bm{G}(\bm{x}_{i}) = \left[G_{1}(\bm{x}_{i}), \ldots,G_{N}(\bm{x}_{i})  \right]$.
Because the operator $\bm{G}$ in equation (\ref{eq::gen_form}) may be discontinuous, an analytic expression for the Jacobian can not be given. However,
by introducing a perturbation vector
$\Delta_{i} = \left[ \delta_{1} , \ldots \delta_{N}  \right]^{T}$, where each element is a scalar value of small magnitude, equation (\ref{eq::gen_form}) can be rewritten in two ways
\begin{equation}
\begin{aligned}
\bm{x}_{i+1} + \Delta_{i+1} &= \bm{G}(\bm{x}_{i}+ \Delta_{i}) \approx   \bm{G}(\bm{x}_{i})+ J\bm{G}(\bm{x}_{i}) \Delta_{i},\\
\bm{x}_{i+1} - \Delta_{i+1} &= \bm{G}(\bm{x}_{i}- \Delta_{i}) \approx   \bm{G}(\bm{x}_{i})- J\bm{G}(\bm{x}_{i}) \Delta_{i}.\\
\end{aligned}
\end{equation}
Here $J\bm{G}(\bm{x}_{i})$ is the Jacobian matrix of $\bm{G}(\bm{x}_{i})$,
which can be approximated as
\begin{equation}
\begin{aligned}
J\bm{G}(\bm{x}_{i}) \Delta_{i} &\approx \bm{G}(\bm{x}_{i} + \Delta_{i}) -   \bm{G}(\bm{x}_{i}), \\
J\bm{G}(\bm{x}_{i}) \Delta_{i} &\approx -\bm{G}(\bm{x}_{i} - \Delta_{i}) +   \bm{G}(\bm{x}_{i}) \, \text{.}
\end{aligned}
\end{equation}
The equations correspond to forward difference and backward difference sche\-mes respectively. Adding these two expressions yields an estimation of 
each column vector of the Jacobian as
\begin{equation}
J\bm{G}_{j}(\bm{x}_{i}) \approx \frac{\bm{G}(\bm{x}_{i} + \Delta_{i}^{j}) - \bm{G}(\bm{x}_{i} - \Delta_{i}^{j})}{2 \delta},
\end{equation}
where $\Delta_{i}^{j} = \delta \bm{e}^{T}_{j}$ and $\bm{e}_{j}$ is the unit vector with unit in the $j$-th element.
From this robust numerical estimation of the Jacobian matrix, any algorithm for calculating the \gls{lle} can be easily employed even for non-smooth applications. Here, the algorithm mentioned in \cite{wolf1986quantifying} is considered.

\subsubsection{Illustrative examples}

\begin{figure}[htp!]
\centering
\includegraphics[scale=0.5]{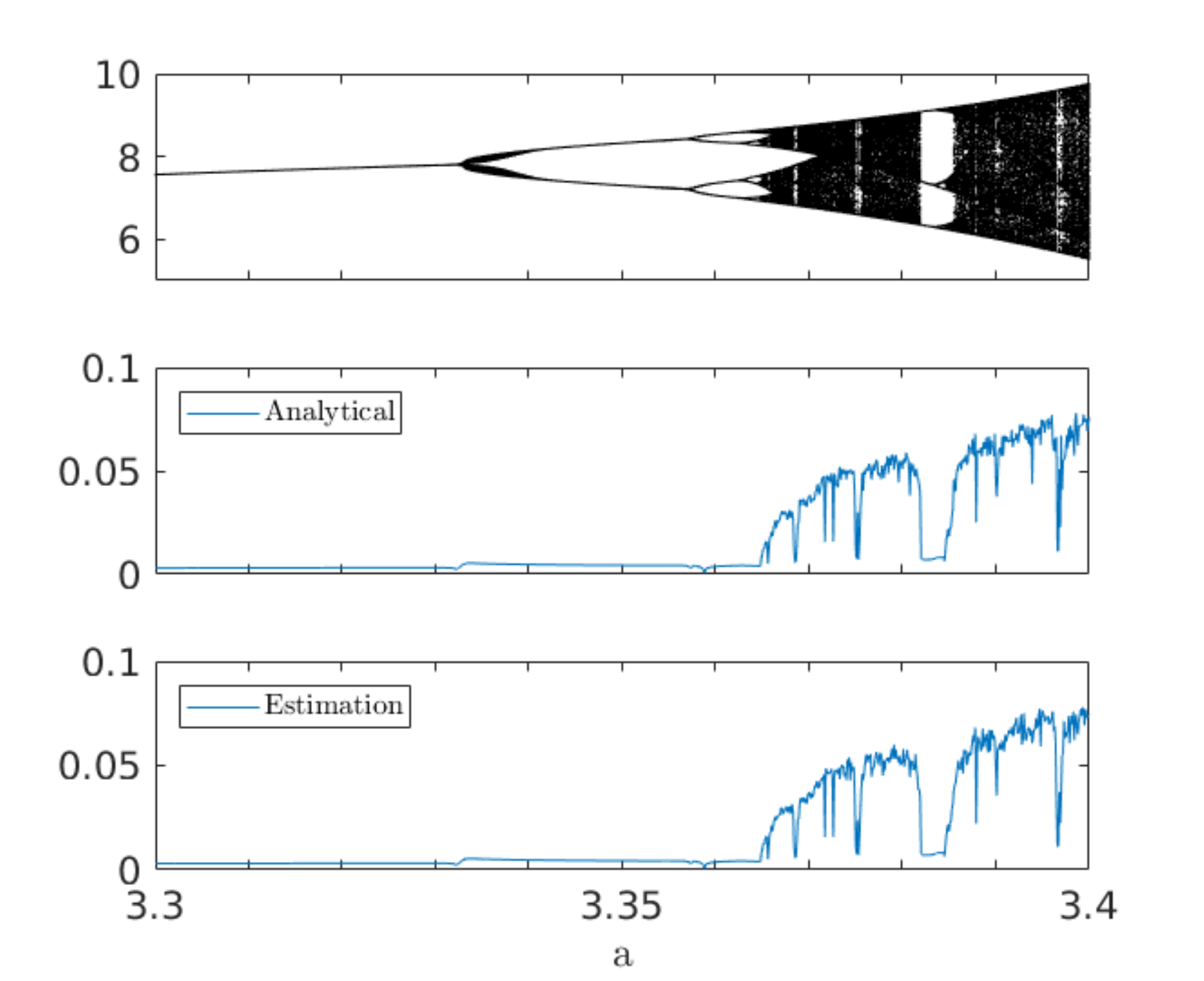}
\caption[LLE estimation technique for continuous problem]{Bifurcation diagram of the problem defined by Eq. (\ref{eq::ProblemFromPaper}) over the parameter $a$ (upper image) analysed with the perspective of the \gls{lle} estimated from the analytic expression of the Jacobian matrix (middle image) and from the numerical approximation of the Jacobian matrix (lower image).}\label{fig:LLEProblemFromPaper}
\end{figure}

The numerical approach is verified using a continuous three-dimensional system for which the analytic Jacobian matrix can be determined. The continuous case suggested by \cite{molaie2013simple} reads
\begin{equation}\label{eq::ProblemFromPaper}
\begin{cases}
\dot{x} & = y, \\
\dot{y} & = z, \\
\dot{z} & = -ax - y - 4z + y^{2} + xy .
\end{cases}
\end{equation}
Here $a$ is a given parameter. The bifurcation diagram of the problem for the domain $a \in [3.3,3.4]$ is plotted in the upper image of Figure \ref{fig:LLEProblemFromPaper}. It can be seen that with an increasing value of the parameter the response gets more chaotic. The estimation of \gls{lle} using the algorithm proposed in \cite{wolf1986quantifying} based on the exact Jacobian matrix of the system, is illustrated in the middle picture. The estimation using the same algorithm but based on the Jacobian matrix approximated by the numerical approach previously introduced in section \ref{sec::EstimationLLE} is plotted in the lowest image in Figure \ref{fig:LLEProblemFromPaper}. 

A perturbation vector $\Delta = 10^{-4} \left[ 1 , 1 , 1 \right]^{T}$ has been employed. It can be observed that the numerical approximation of the Jacobian matrix performs well. The results provided by the analytical expression are accurately reproduced. It can also be seen that \gls{lle} mirrors the chaotic behavior of the bifurcation diagram well. 

\begin{figure}[htp!]
\centering
\includegraphics[scale=0.5]{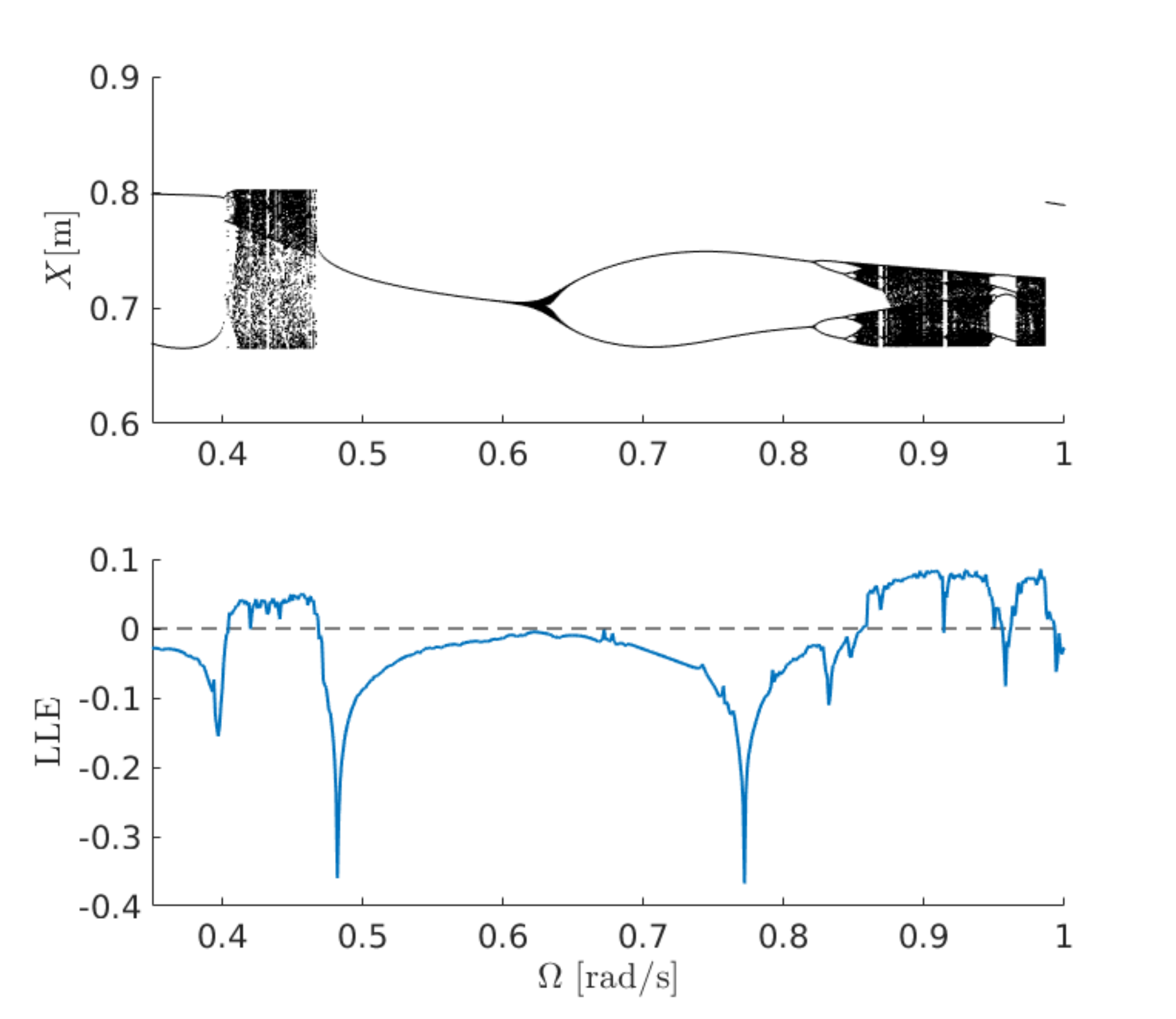} 
\caption[\gls{lle} estimation technique for discontinuous problem]{Description of bifurcations by positive values of the \gls{lle}}
\label{fig:LLEOverBifurcationMyProblem}       
\end{figure}

Let consider the discontinuous system of ordinary differential equations defined by equation (\ref{eq::dynamic_system}). Here LLE values can only be evaluated from the numerical approximation. Using a perturbation vector $\Delta = 10^{-4} \left[ 1 , 1 , 1 \right]^{T}$, the bifurcation diagram and the respective LLE values for this dynamic problem are plotted over the angular frequency in Figure \ref{fig:LLEOverBifurcationMyProblem}. It can be seen that the positive values of the \gls{lle} over the parametric domain correspond well with the bifurcation graph. Thus the \gls{lle} appears again as a pertinent instability indicator.

\subsubsection{Largest Lyapunov exponent as an instability indicator}

In order to accurately determine regular or chaotic motion of the system, it is not required to quantitatively estimate the largest Lyapunov exponent, but it seems rather more promising to use its non-negativity or negativity as a binary indicator.

\begin{figure}[htp!]
\centering
\subfloat[\gls{lle} values over parametric space]{\label{fig:LLEExample1_plot}
\includegraphics[width=0.48\textwidth]{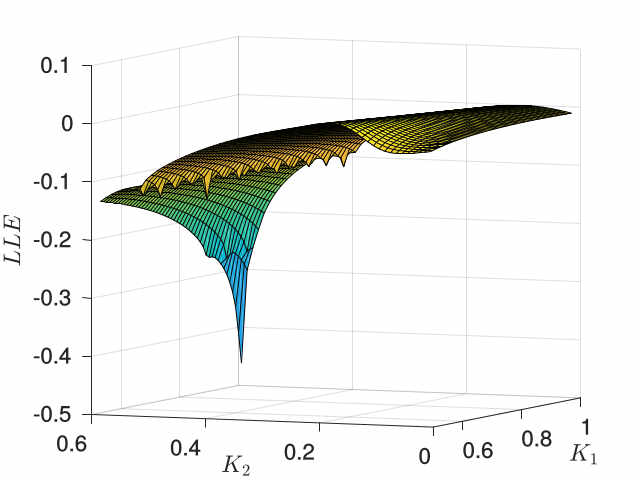}}
\subfloat[Indicator $\mathcal{C}_{MoB}$ map (red: $\mathcal{C}_{MoB}=1$, gray: $\mathcal{C}_{MoB}=0$)]{\label{fig:LLEExample1_redgray}
\includegraphics[width=0.48\textwidth]{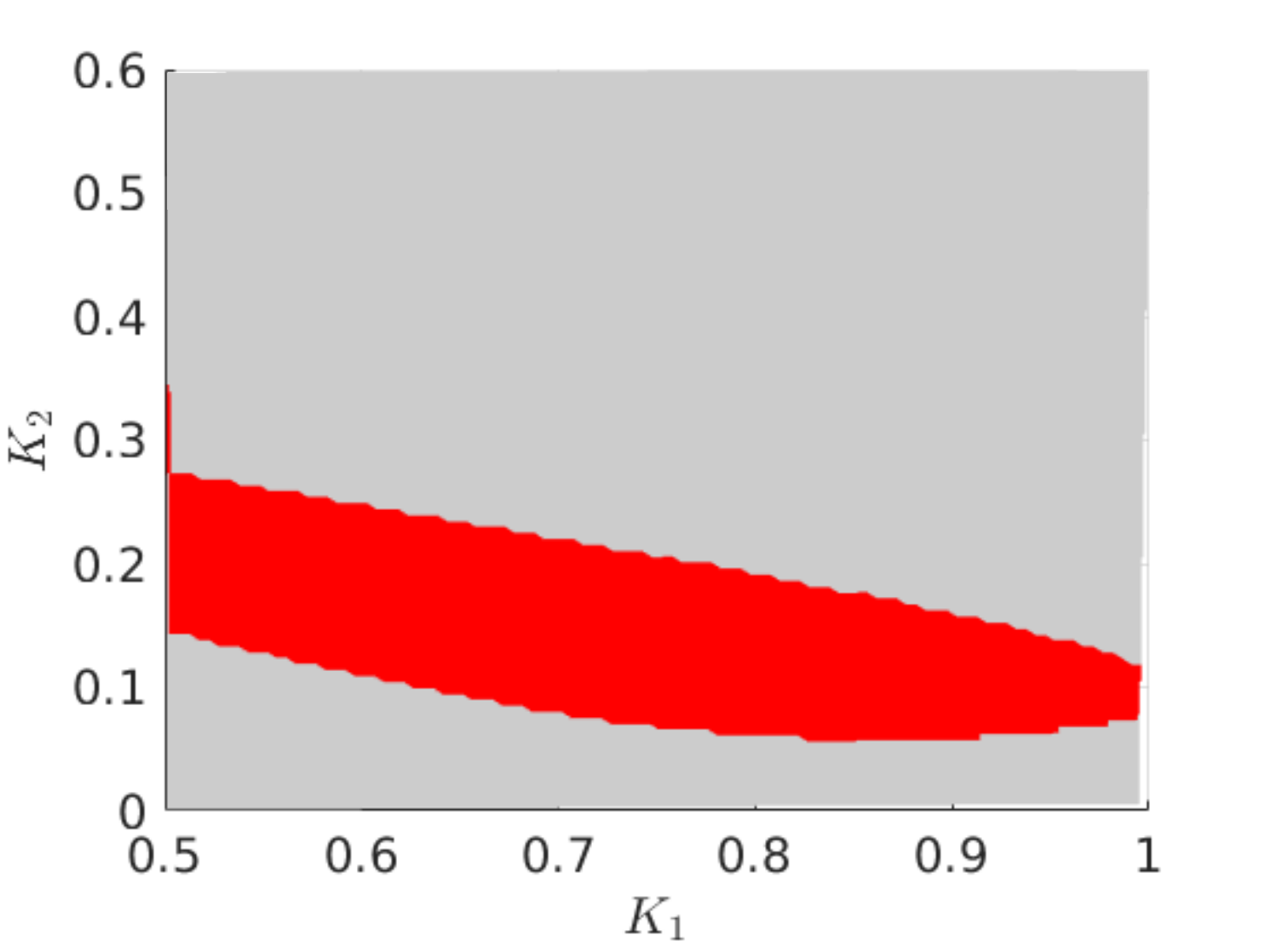}}
\caption[Example of LLE values over the spring stiffnesses]{Example of \gls{lle} analysis over the spring stiffnesses ($K_{1}$ in $\text{N}.\text{m}^{-3}$, $K_{2}$ in $\text{N}.\text{m}^{-1}$)}\label{fig:LLEExample1}
\end{figure}

From $\text{LLE}_{MoB}(\bm{x})$ which provides the \gls{lle} value for the \gls{mob} system with input $\bm{x}$ over the input parametric space, an instability indicator $\mathcal{C}_{MoB}(x)$  can be defined as 
\begin{equation}\label{eq::ClassifierMotion}
\mathcal{C}_{MoB}(x) = \begin{cases}
1, & \text{if} \, \, \text{LLE}_{MoB}(\bm{x}) \geq 0, \\
0, & \text{if} \, \, \text{LLE}_{MoB}(\bm{x}) < 0 .
\end{cases}
\end{equation} 
For sake of illustration, the set of \gls{lle} values represented in Figure \ref{fig:LLEExample1_plot} is analyzed through this instability indicator, which is plotted in Figure \ref{fig:LLEExample1_redgray}. The stable behavior associated with values of the indicator $\mathcal{C}_{MoB}=0$ is indicated in gray, whereas the parametric domain corresponding with chaotic behavior i.e. $\mathcal{C}_{MoB}=1$ is highlighted in red. It can be seen that this \gls{lle} surface provides a valuable information which could allow to optimize the system by preventing instabilities.

Based on this proficient \gls{qoi}, the dynamic system could be efficiently analyzed through a surrogate model to evaluate the critical values at low cost. The goal of the surrogate model is to be able to provide an accurate approximation of this surface with only few observations, originating either from physical or numerical experiments. This surrogate model should be the best representation of the surface with respect to the available information, and provide guidance for defining further experiments to be conducted. 

\section{Construction of surrogate model}
\label{sec:surrogate}

The idea behind surrogate models is only summarized here. Given a mapping $\mathcal{M} \, : \mathbb{X} \rightarrow \mathbb{Y}$ between an input $\bm{x} \in \mathbb{X} \subset \mathbb{R}^{n}$ and some uni-variate output $\bm{y} \in \mathbb{Y} \subset \mathbb{R}$. From a set of $m$ observations $\mathcal{D} = \lbrace \left( \bm{x}^{(i)}, \,\bm{y}^{(i)} \right), \, i=1, \, \ldots  , \, m  \rbrace$, global surrogate modeling aims to statistically approximate the output on the whole input space, i.e. to approximate the  statistical relationship behind this mapping. The samples in the continuous input space are called design of experiments and are denoted by the set $\mathcal{X} = \lbrace \bm{x}^{(i)}, \, i=1, \, \ldots  , \, m  \rbrace$. For uni-variate output the mapping evaluations are gathered in the vector $\bm{y} = \lbrace y_{i}, \, i=1, \, \ldots  , \, m  \rbrace$. The statistical approximation on the whole input space will be given by a computed surrogate model denoted by $\tilde{\mathcal{M} }$. Among the different existing surrogate modeling approaches, kriging is investigated here.

\subsection{Kriging}
 Kriging was proposed by a mining engineer called Krige for geostatistics study \citep{krige1951statistical}. The mathematical foundation for the approach has been developed by \cite{matheron1963principles}, which also established its name. For an overview the reader is referred to \cite{kleijnen2009kriging} and \cite{kleijnen2017regression}. It can thus be outlined that the approach can be described from a regression perspective as a statistical extension of deterministic regression \cite{stein1991universal}, or from a bayesian or machine learning perspective as an update from prior information based on a conditional distribution \cite{williams1996gaussian}.

 \subsubsection{Ordinary kriging metamodel}

\begin{figure}[htp!]
    \centering
\includegraphics[width=0.48\textwidth]{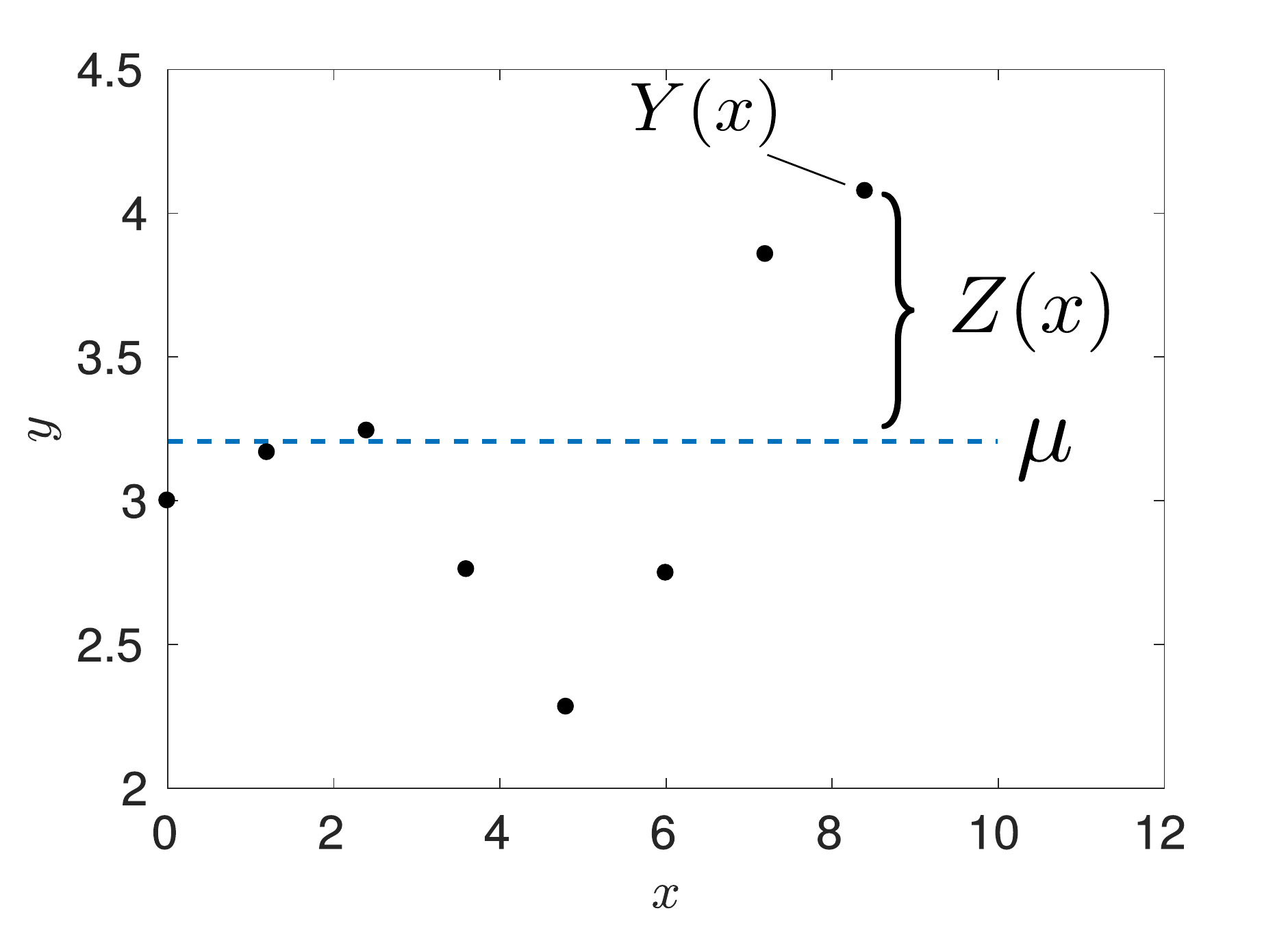}
    \caption{Scheme of an ordinary kriging metamodel $Y(x)$ as the superposition of a constant mean $\mu$ of the samples, shown by black circles, and an uncertain deviation described by the Gaussian process $Z(x)$.}
    \label{fig:scheme_kriging}
\end{figure}

 Using ordinary kriging \cite{kleijnen2017regression}, the mapping between input and output data is approximated by a combination of a global contribution characterized by $\mu$, an unknown constant describing the mean contribution, and localized departures denoted by $Z(\bm{x})$
 \begin{equation}\label{eq::Kriging}
Y(\bm{x}) =  \mu + Z(\bm{x})  \text{.}
\end{equation}
In details, $Z(\bm{x})$ is a realization of a stochastic process with zero mean and a variance $\sigma^{2}$ accounting for local deviations. The covariance matrix of $Z(\bm{x})$ is given as
 \begin{equation}
     Cov [Z(\bm{x}^{(i)}), Z(\bm{x}^{(j)})] = \sigma^{2} R(\bm{x}^{(i)},\bm{x}^{(j)}, \bm{\theta}) \, \text{.}
 \end{equation}
 Here $R(\bm{x}^{(i)},\bm{x}^{(j)}, \bm{\theta})$ is the correlation function between two points of the input set. $\bm{\theta}$ is called hyperparameter and contains the unknown correlation parameters between two points. It can thus be seen, as illustrated in Figure \ref{fig:scheme_kriging}, that the ordinary kriging model depends on two main contributions, a regression part which is here only a mean contribution, and a stationary Gaussian process based on the assumption of an auto-correlation function, which governs the deviation from that mean. 

  \subsubsection{Prior information}
  
Kriging model is based on some observations, as prior information within a Bayesian approach \cite{handcock1993bayesian}. It could also be considered as a training stage of the surrogate model. Considered in ordinary kriging as a constant, the prior value of the mean $\hat{\mu}$ can be obtained from the available observation $\bm{y}$ using a generalized least-square estimate as
\begin{equation}
\hat{\mu} = (\bm{1}^{T} \bm{R}^{-1} \bm{1})^{-1} \bm{1}^{T} \bm{R}^{-1} \bm{y}
\end{equation}

 \noindent with $\bm{1}$ a vector containing only components with scalar value $1$ and $\bm{R}$ the correlation matrix. It can be noticed that this prior evaluation depends on the chosen correlation function. The evaluation of the standard deviation from the prior information is denoted by $\hat{\sigma}$. Then, the variance of the prior information reads
\begin{equation}
\hat{\sigma}^{2} = \frac{1}{m} \left( \bm{y} - \bm{1} \hat{\mu} \right)^{T} \bm{R}^{-1} \left( \bm{y} - \bm{1} \hat{\mu} \right) \text{,}
\end{equation}
 which, similarly to the mean, depends on the chosen correlation matrix.
 
\subsubsection{Auto-correlation function and hyperparameter evaluation}

The realization of the Gaussian process depends on its characteristics, namely its mean, its variance and its auto-correlation. 

 The auto-correlation function, commonly referred as correlation function for sake of simplicity, is usually assumed by the user. A common choice is the Mat\'{e}rn 3/2 correlation function \citep{matern1960spatial}, defined as
\begin{equation}
R (\bm{x} - \bm{x'}, \bm{l})  = \prod_{i=1}^{n} \left( 1 + \dfrac{\sqrt{3} \abs{x_{i} - x_{i}'} }{l_{i}} \right) \, \exp \left(-\dfrac{\sqrt{3} \abs{x_{i} - x_{i}'} }{l_{i}}  \right) \, \text{,}
\end{equation}
in which a different hyperparameter $l_i$ can be used for each dimension $n$ of the design of experiments. Due to the lack of knowledge about the observations this correlation structure is arbitrary chosen here.

The hyperparameters need to be determined before constructing the surrogate model. This can for example be done by utilizing the maximum likelihood estimate of the hyperparameter value.
This method yields an auxiliary optimization problem \cite{dubourg2013metamodel} for estimating $\bm{\theta}$ of the form 
\begin{equation}\label{eq::minimization_hyperparameters}
\hat{\bm{\theta}} =   \arg \, \min_{\bm{\theta}^{\star}}  \psi (\bm{\theta}^{\star}) \, \text{,}
\end{equation}
in which $\psi$ is the reduced likelihood defined for ordinary kriging as
\begin{equation}
\begin{aligned}
\psi (\bm{\theta}) = \hat{\sigma}^{2} (\bm{\theta}) [\det \bm{R} (\bm{\theta})]^{1/m},
\label{eq:optimization}
\end{aligned}
\end{equation}
with $\det$ denoting the determinant operator.

 Since there is no analytic solution for the optimization problem defined by Equation (\ref{eq:optimization}), it is necessary to use numerical optimization tools. Bouhlel and Martin mention this step as being the most challenging for the construction of surrogate models with kriging because of the multimodality of the likelihood function \cite{bouhlel2019gradient}. In this work, a hybridized particle swarm optimization is employed similar to the method suggested by \cite{toal2011development}.

\subsubsection{Surrogate model evaluation}

Consider the unobserved value $Y_{0} \equiv Y(\bm{x}^{(0)})$ with $x^{(0)} \in \mathbb{X}$, which needs to be predicted. 
When utilizing the best linear unbiased predictor for the Gaussian random variate $\hat{Y}_{0}$,
the mean of the unobserved value yields 
\begin{equation}
\mu_{\hat{Y}_{0}}  =  \hat{\mu} + \bm{r}_{0}^{T} \bm{R}^{-1} (\bm{y}- \bm{1} \hat{\mu}).
\end{equation}
 $\bm{r}_{0}$ describes the cross-correlations between the new point $\bm{x}^{(0)}$ and each available observation with 
\begin{equation}
r_{0 \, i} = R(\bm{x}^{(0)} - \bm{x}^{(i)}, \bm{\theta}) \, \qquad i=1, \, \ldots \, , m.
\end{equation}
Furthermore the variance of the unobserved value can be estimated as
\begin{equation}
\sigma_{\hat{Y}_{0}}^{2}  = \sigma^{2} \left( 1 - \bm{r}_{0}^{T} \bm{R}^{-1} \bm{r}_{0} + \bm{u}_{0}^{T} \left( \bm{1}^{T} \bm{R}^{-1} \bm{1}\right)^{-1} \bm{u}_{0}\right),
\end{equation}
with
\begin{equation}
\bm{u}_{0} = \bm{1}^{T} \bm{R}^{-1} \bm{r}_{0} - \bm{1} \, \text{.}
\end{equation}
For proof and further details the reader shall be referred to \cite{santner2013design}.

\subsection{Adaptive sampling}

To optimize the interest of the computations, the observation points which define the experimental design are chosen with respect to an adaptive sampling scheme. Starting from a set of initial data $\mathcal{D}_{ini} = \lbrace \left( \bm{x}^{(i)}, \,\bm{y}^{(i)} \right), \, i=1, \, \ldots  , \, m  \rbrace$ a surrogate model $\tilde{\mathcal{M}}$ is constructed with kriging based on this knowledge. By solving some auxiliary optimization problems new points are successively added to the dataset $\mathcal{D}$ until a convergence criterion is reached. The general optimization problem reads
\begin{equation}\label{eq::}
\bm{x}^{(m+1)} = \arg \, \min_{\bm{x}^{\star} \in \mathbb{X}} \, \text{Score} \left( \bm{x}^{\star} \right) \, \text{.}
\end{equation}
Here a score function is evaluated which describes generally a tradeoff between an exploration and an exploitation component. Using the exploration term, the domain is globally examined in order to detect unknown regions of interest. The exploitation aims to generate data points in the pre-identified regions of interest to reduce the prediction error locally. Different techniques have been proposed in the literature such as \citep{jones1998efficient,turner2007multidimensional,singh2013balanced}.

An external variance-based technique has recently been proposed in \cite{liu2017adaptive}. \gls{mepe} is a continuous optimization involving a switch strategy between exploitation and exploration. The bias error is estimated by a continuously approximated leave-one-out cross validation (LOOCV) error $\hat{e}(\bm{x})$. The continuous expected prediction error to be maximized is given by
\begin{equation}\label{eq::MEPE_eq}
EPE(\bm{x}) = \alpha  \hat{e}(\bm{x}) + (1-\alpha) \hat{\sigma}_{\hat{Y}}^{2},
\end{equation}
where $\hat{e}(\bm{x})$ is the value of $\hat{e}(\bm{x}^{(i)})$ when $\bm{x}$ is located in the Voronoi cell of point $\bm{x}^{(i)}$. Otherwise its value is zero. In order to fasten the computation an approximation of $\hat{e}(\bm{x}^{(i)})$ can be used \cite{liu2017adaptive,sundararajan2000predictive}.  A balance factor $\alpha$ is introduced to adjust the exploitative bias term and the exploratory variance term by switching adaptively between the two components depending on the estimation quality of the bias term. The computation of the balance factor for the calculation of the sample point $\bm{x}^{(m+q-1)}$ reads
\begin{equation}\label{eq::alphaFactor}
\begin{aligned}
\alpha = \begin{cases} 0.5, & \text{if} \, q=1 \\
0.99 \min \left[ 0.5 \dfrac{ e_{true}^{2}(\bm{x}^{(m+q-1)}) }{\hat{e}(\bm{x}^{(m+q-1)})}, 1 \right] & \text{if} \, q>1.
\end{cases}
\end{aligned}
\end{equation} 
Here, $e_{true}$ represents the error between the surrogate model and the actual observation for this sample point.
The algorithm is summarized in Box \ref{alg::MEPE}.
\begin{kasten}[htbp!]
\begin{Algorithmus}
\begin{itemize}
\itemsep1em 
\item Given a design of experiments $\mathcal{X} = \lbrace \bm{x}^{(1)}, \ldots , \bm{x}^{(m)} \rbrace$. 
\item[] While the adaptive sampling stopping criterion is not satisfied. \\
Do: 
\begin{itemize}
\item[] Update the balance factor $\alpha$ with equation (\ref{eq::alphaFactor}).
\item[] Obtain the LOOCV error at each sample point $\hat{e}(\bm{x}^{(i)})$.
\item[] Obtain a new point by maximizing the $EPE$-criterion of equation (\ref{eq::MEPE_eq}) over the input 
\item[] Update the design of experiments with the new found point and its response.
\end{itemize}
end
\end{itemize}
\end{Algorithmus}
\captionof{kasten}{Algorithm for MEPE}\label{alg::MEPE}
\end{kasten}

The second scheme considered here is Expected Improvement for Global Fit (EIGF) \cite{lam2008sequential}, which is also a continuous optimization scheme, but with a fixed balance between exploitation and exploration. It is based on the maximization of the expected improvement defined as
\begin{equation}
EI(x) = (\mu_{\hat{Y}}(x) - y(x_{\ast}) )^{2} + \sigma_{\hat{Y}}^{2}(x)
\end{equation}
where $x_{\ast}$ is the closest sample point to the candidate $x$. The first term, which represents the exploitation contribution, is large when the surrogate model $\hat{Y}(x)$ differs substantially from the response at the nearest point. The second term provides global exploration, it is large when the surrogate model exhibits higher lack of knowledge.

For classification problems, a dedicated adaptive scheme called \gls{mivor} has recently been proposed \cite{Class_Krig_sub,Jan_Master_thesis}. That scheme is based on the exploration of the parametric space by Voronoi tessellation. A score is given to all cells to identify the one cell corresponding to the largest lack of knowledge near to a decision region between two classes and so to pertinently add a new observation point. A random switching strategy is used to balance exploration and exploitation during the successive iterations. This allows to reach a compromise between the exploration of the whole parametric space and the addition in some local areas near to the class boundaries which require to be accurately described. The scheme has been described in detail, investigated for various applications and compared with other adaptive schemes in \cite{Class_Krig_sub}.

\section{Surrogate model for dynamic  analysis}
\label{sec:Kriging_dyn}

The surrogate strategy is tested on different applications. The problem parameters are summarized in Table \ref{tab:parameters}. One-, two- and three-dimensional parametric spaces are considered.
\begin{table}[htbp!]
\centering
\resizebox{\textwidth}{!}{  
\begin{tabular}{|c|c|c|c|c|c|c|}
\cline{2-7}
\multicolumn{1}{c|}{} & \multicolumn{6}{c|}{Problem} \\
 \cline{2-7}
\multicolumn{1}{c|}{} &  $\mathcal{P}_0$ & $\mathcal{P}_1$ & $\mathcal{P}_2$ & $\mathcal{P}_3$ & $\mathcal{P}_4$ & $\mathcal{P}_5$  \\
\hline
$M$  & \multicolumn{6}{c|}{$1 \, \text{kg}$} \\
$V_0$  & \multicolumn{6}{c|}{$0.1 \, \text{m}.\text{s}^{-1}$} \\
$D$  & \multicolumn{6}{c|}{$0.0 \, \text{N}.\text{s}.\text{m}^{-1}$} \\
$\mu_s$  & \multicolumn{6}{c|}{$0.3$} \\
$V_s$  & \multicolumn{6}{c|}{$0.1 \, \text{m}.\text{s}^{-1}$} \\
$U_0$  & \multicolumn{6}{c|}{$0.1 \, \text{N}$} \\
$N_0$  & \multicolumn{6}{c|}{$1.0 \, \text{N}$} \\
$\sigma_0$  & \multicolumn{6}{c|}{$100.0 \, \text{N}.\text{m}^{-1}$} \\
$\sigma_1$  & \multicolumn{6}{c|}{$10.0 \, \text{N}.\text{s}.\text{m}^{-1}$} \\
$\sigma_2$  & \multicolumn{6}{c|}{$0.1 \, \text{N}.\text{s}.\text{m}^{-1}$} \\
\hline
$\Omega$ (in $\text{rad}.\text{s}^{-1}$)  & $0.6 $ & $ \left[ 0.2,1.0\right]$ & $0.6 $ & $0.6$ & $0.7 $ & $\left[ 0.6,0.9\right]$ \\
\hline
$K_1$ (in $\text{N}.\text{m}^{-3}$) & $\left[ 0.5,1.0\right]$ & $ 1.0$ & $\left[ 0.5,1.0\right]$ & $1.0$ & $\left[ 0.5,1.0\right]$ & $1.0$ \\
\hline
$K_2$ (in $\text{N}.\text{m}^{-1}$) & $\left[ 0.0,0.6\right]$ & $ 0.0 $  & $\left[ 0.0,0.5\right]$ & $\left[ 0.0,0.5\right]$ & $\left[ 0.0,0.6\right]$ & $\left[ 0.0,0.5\right]$ \\
\hline
$\mu_k$  & $0.15$ & $0.15$ & $0.15$ & $\left[ 0.08,0.18\right]$ & 0.15 & $\left[ 0.10,0.15\right]$ \\
\hline
    \end{tabular}}
    \caption{Problem parameters}
    \label{tab:parameters}
\end{table}

The performances for both \gls{qoi}s are analyzed with respect to a reference solution computed with $m_{ref}$ evaluation points $\lbrace x_{ref,i}, \, i = 1,\dots, m_{ref} \rbrace$. 5000 points per dimension are considered for the applications. Four indicators are considered to evaluate the error between the metamodel $\hat{Y}$ and the reference response surface $y_{ref}$. The \gls{mae} is estimated considering the reference points as 
\begin{equation}
\text{MAE} = \dfrac{1}{m_{ref}} \sum_{i=1}^{m_{ref}} \abs{y_{ref}(x_{ref,i}) - \hat{Y}(x_{ref,i})}.
\label{eq:MAE}
\end{equation}
With the same perspective, the \gls{rmse} is estimated as
\begin{equation}
\text{RMSE} = \sqrt{\dfrac{1}{m_{ref}} \sum_{i=1}^{m_{ref}} \left(y_{ref}(x_{ref,i}) - \hat{Y}(x_{ref,i})\right)^{2}}.
\end{equation}
The \gls{rmae} is given by
\begin{equation}
\text{RMAE} = \dfrac{\max\limits_{i \in[1,m_{ref}]} \left( \abs{y_{ref}(x_{ref,i}) - \hat{Y}(x_{ref,i})}\right)}{\sigma_{y_{ref}}},
\end{equation}
with $\sigma_{y_{ref}}$ the standard deviation of the reference solution. Finally the $\text{R}^{2}$ score reads
\begin{equation}
\text{R}^{2} = 1 - \dfrac{\sum_{i=1}^{m_{ref}} \left(y_{ref}(x_{ref,i}) - \hat{Y}(x_{ref,i})\right)^{2}}{\sum_{i=1}^{m_{ref}} \left(y_{ref}(x_{ref,i}) - \overline{y}_{ref}\right)^{2}},
\end{equation}
where $\overline{y}_{ref}$ denotes the mean of the reference response values.
The parametric input space is normalized in each dimension in order to avoid numerical issues. Let $x_{i}^{u}$ and $x_{i}^{l}$ be the upper and lower limit of the dimension $i$ respectively. The normalized input  $x_{i}$ is then given by
\begin{equation}
    \bar{x}_{i} = \frac{x_{i} - x_{i}^{l}}{x_{i}^{u} - x_{i}^{l}}.
\end{equation}{}
The normalization ensures that $0 \leq \bar{x}_{i} \leq 1$ and is denoted by a bar.


\subsection{Surrogate model for sticking time: Problem $\mathcal{P}_{0}$}

First, the ability of the surrogate model to provide an estimation for the sticking time is investigated. The parametric domain of the problem $\mathcal{P}_{0}$ is given by the two spring stiffnesses with $K_{1} \in  [0.5 , 1.0]\, \text{N}.\text{m}^{-3}$
and $K_{2} \in [0.0,0.6] \, \text{N}.\text{m}^{-1}$. 
Other parameter values are considered to be deterministic, see Table \ref{tab:parameters}. 

\begin{figure}[htp!]
\centering
\subfloat[Reference solution]{\label{fig:STickingKSeta06Plot} \includegraphics[width=0.53\textwidth]{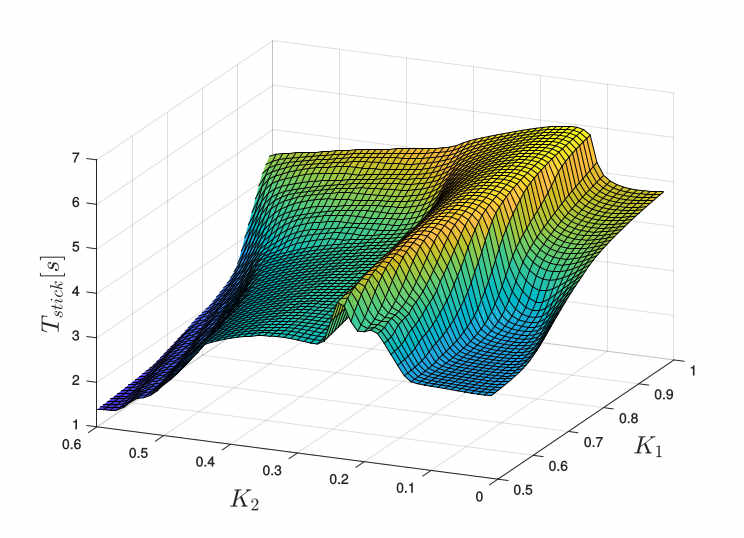}}
\subfloat[Contour of absolute error (in s) for the initial surrogate model (20 samples with \gls{tplhd})]{\label{fig:STickingKSeta06Error}
\includegraphics[width=0.45\textwidth]{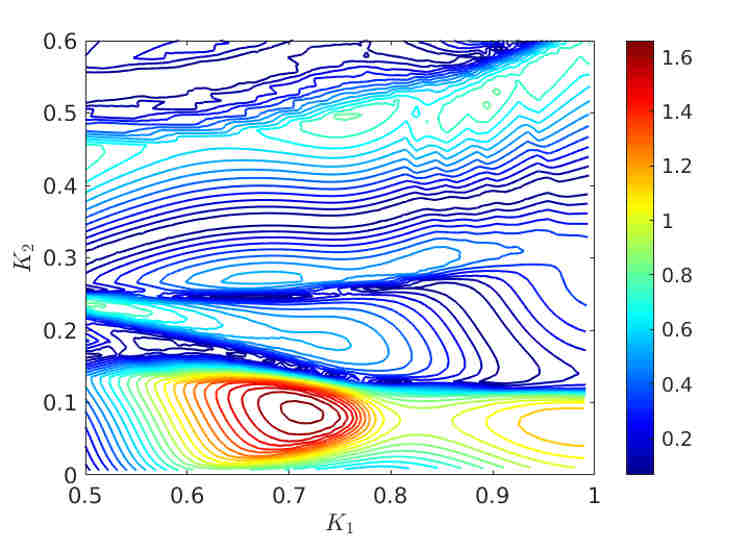}}
\caption[Sticking time and absolute error (in s) of the evaluation of the sticking time]{Sticking time full response surface (Problem $\mathcal{P}_{0}$) and initial absolute error  of the metamodel. $K_{1}$ is given in $\text{N}.\text{m}^{-3}$, $K_{2}$ in $\text{N}.\text{m}^{-1}$.}\label{fig:STickingKSeta06}
\end{figure}

The reference response surface for the sticking time over the input domain is plotted in Figure \ref{fig:STickingKSeta06Plot}. Initially 20 samples are spread across the domain with \gls{tplhd}  \cite{viana2010algorithm} leading to a metamodel with an absolute error distributed as shown in Figure \ref{fig:STickingKSeta06Error}.
It can be noticed that the spread of the sticking time in the parametric domain $\lbrace K_{1}, K_{2}\rbrace$ is around 5 seconds and the highest initial absolute error is around 1.6 seconds. 

\subsubsection{Adaptive kriging strategy}

From the initial \gls{tplhd} model with 20 samples, surrogate models utilizing 60 samples are built considering the two alternative sampling techniques previously introduced, i.e. \acrfull{mepe} and \acrfull{eigf}. Furthermore a direct TPLHD model with 60 samples is generated. In order to avoid numerical outliers of the stochastic optimization process each strategy is executed twenty times, so that the performances can be examined as mean performance of these realizations.

The error measures for the different surrogate models are listed in Table \ref{table::StickingTime}. The two adaptive schemes provide much more proficient metamodels than the initial and final one-shot metamodels based on \gls{tplhd}. Among them EIGF appears more proficient than \gls{mepe}. 
\begin{table}[htbp!]
\begin{center}
\begin{tabular}{l|l c c c c} 
\hline
\\
& Method & MAE [s] & RMAE & RMSE [s] & R$^{2}$  \\ 
\\
\hline
\hline
\\
\multirow{1}{*}{\shortstack[l]{Metamodel \\ with 20 samples}} & TPLHD & 0.31 & 1.29 & 0.46 & 0.88 \\ \\ 
\hline 
\\
\multirow{4}{*}{\shortstack[l]{Metamodel \\ with 60 samples}} &TPLHD & 0.28 & 1.27 & 0.38 & 0.93 \\
& & & & & \\
&EIGF &  0.16 & 1.00 & 0.22 & 0.97 \\
&MEPE &0.17 & 1.09 & 0.27 & 0.96 \\
\\
\end{tabular}
\end{center}
\caption[Error measures for the sticking time metamodel 60 samples]{Error measures for the sticking time metamodels (Problem $\mathcal{P}_{0}$) with 60 samples.}\label{table::StickingTime}
\end{table}

\begin{figure}[htbp!]
\centering
\subfloat[MAE error in s]{\label{fig:TStickKSeta06YesMAEerror}
\includegraphics[width = 0.49\textwidth]{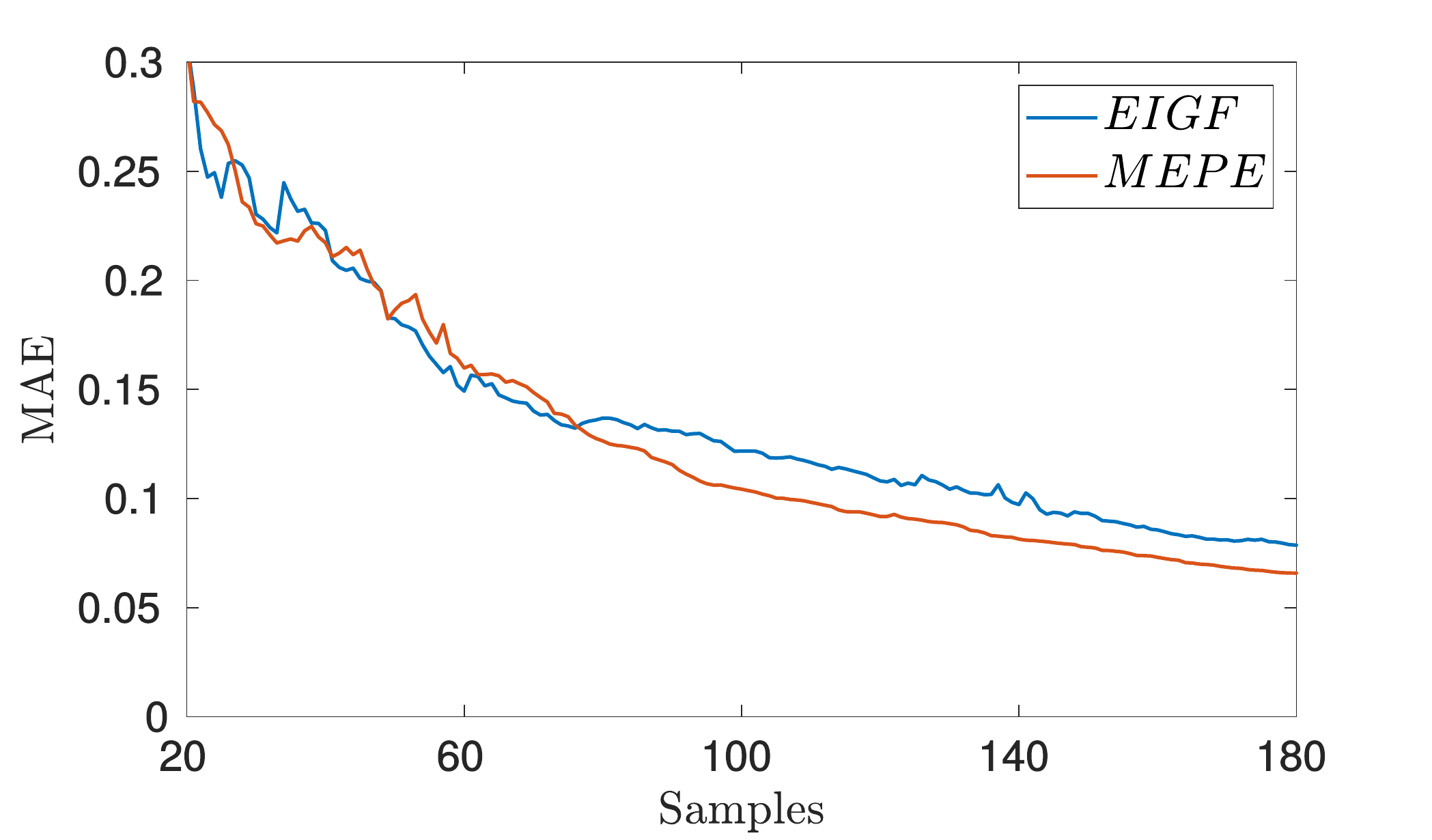}}
\subfloat[RMSE error in s]{\label{fig:TStickKSeta06YesRMSEerror}
\includegraphics[width = 0.49\textwidth]{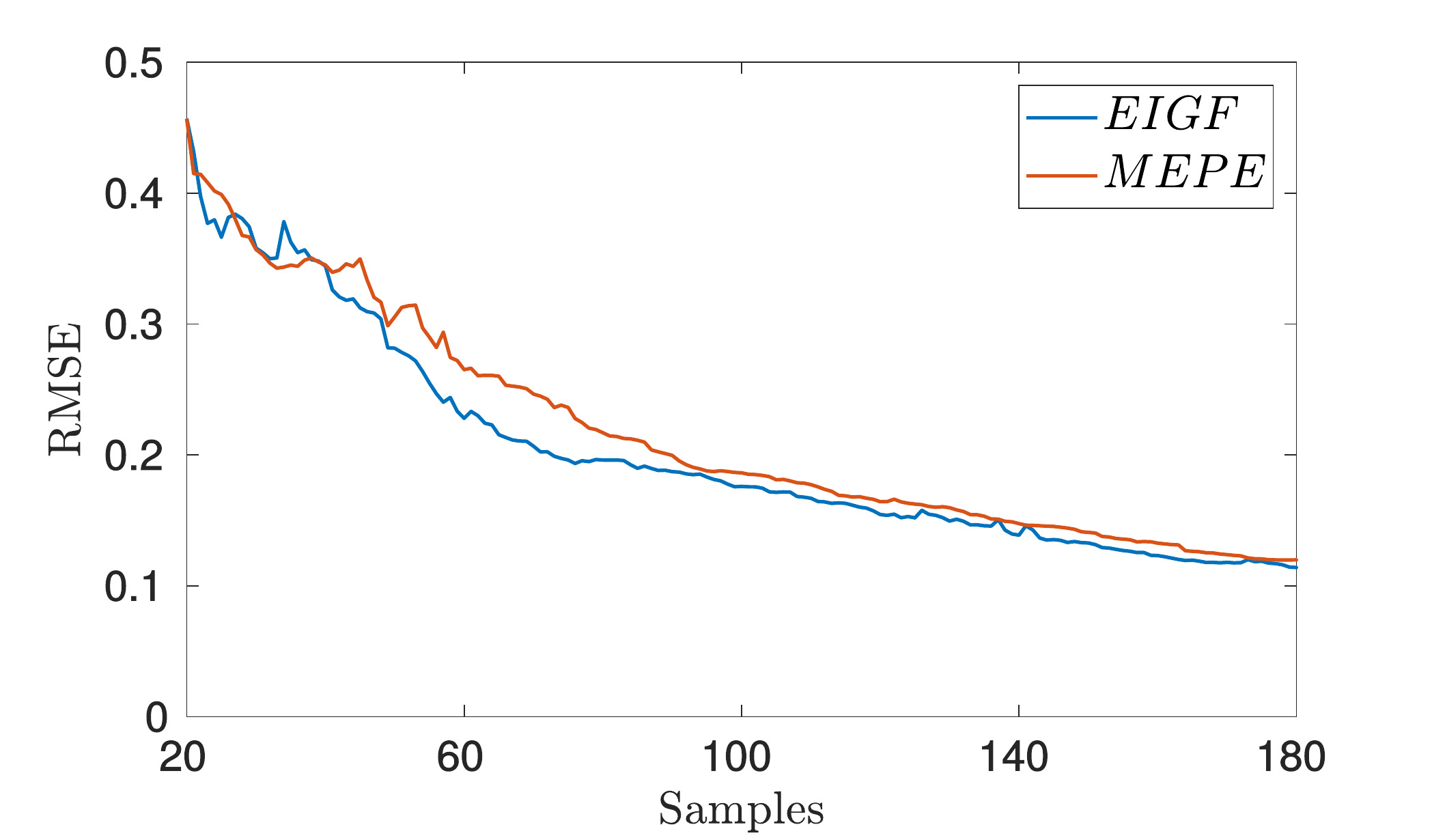}}
\caption[MAE error over the sample size for the sticking time problem]{Evolution of the errors with the sample size for the two adaptive sampling techniques for the sticking time metamodels (Problem $\mathcal{P}_{0}$) until 150 samples.}\label{fig::KSetaConvergence}
\end{figure}

In Figure~\ref{fig::KSetaConvergence} the convergence of MAE error and RMSE error with the number of samples is plotted for the two adaptive methods. As before, all results are averaged on twenty realizations. It seems that MEPE performs better as shown in Figure \ref{fig:TStickKSeta06YesMAEerror}. However, by considering the convergence in terms of RMSE error, both methods show a similar convergence behaviour, see Figure \ref{fig:TStickKSeta06YesRMSEerror}.

\begin{figure}[htp!]
\centering

\subfloat[EIGF - 150 samples]{\label{fig:TStickKSeta06YesEIGF}
\includegraphics[width = 0.49\textwidth]{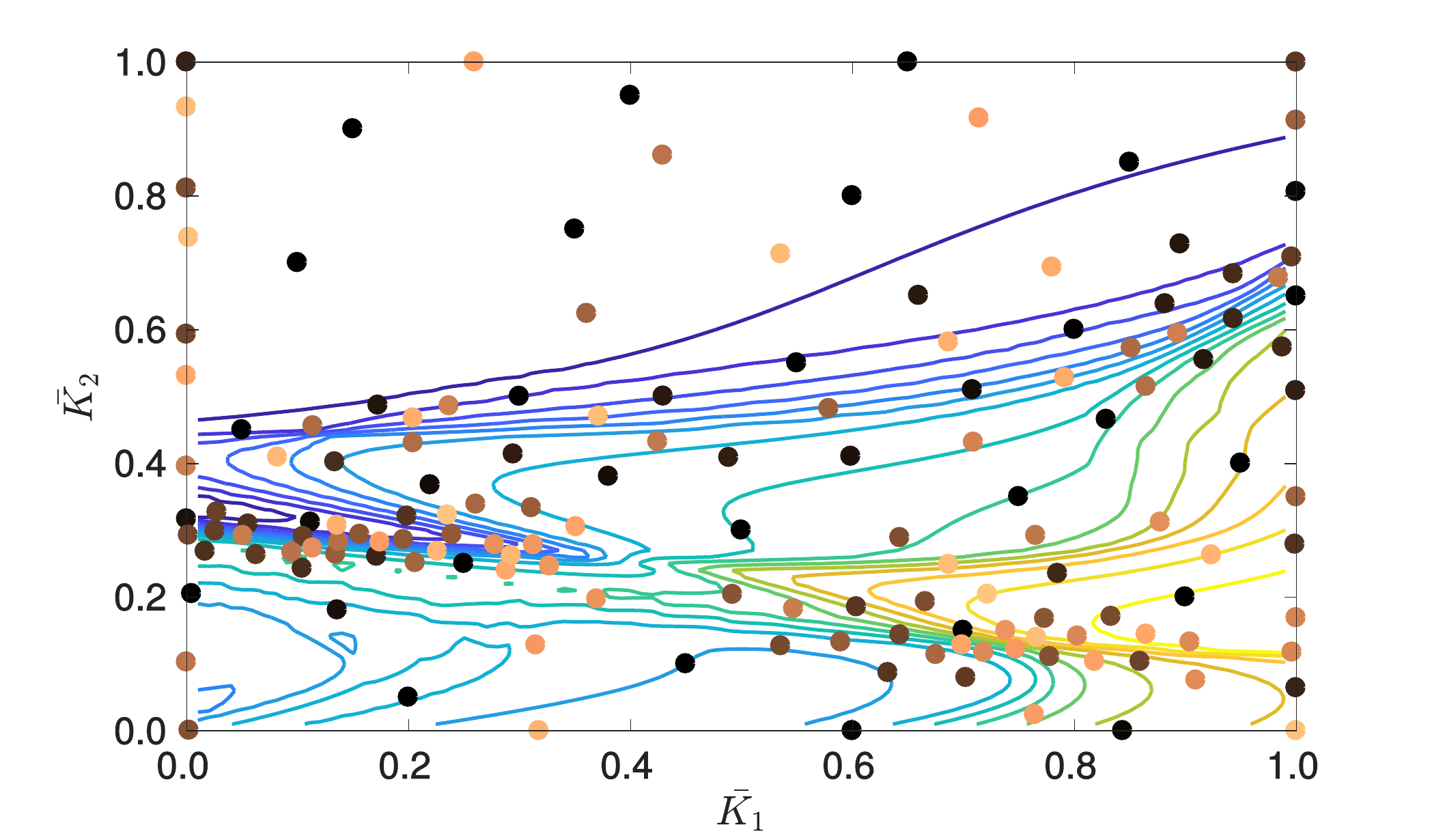}}
\subfloat[MEPE - 150 samples]{\label{fig:TStickKSeta06YesMEPE}
\includegraphics[width = 0.49\textwidth]{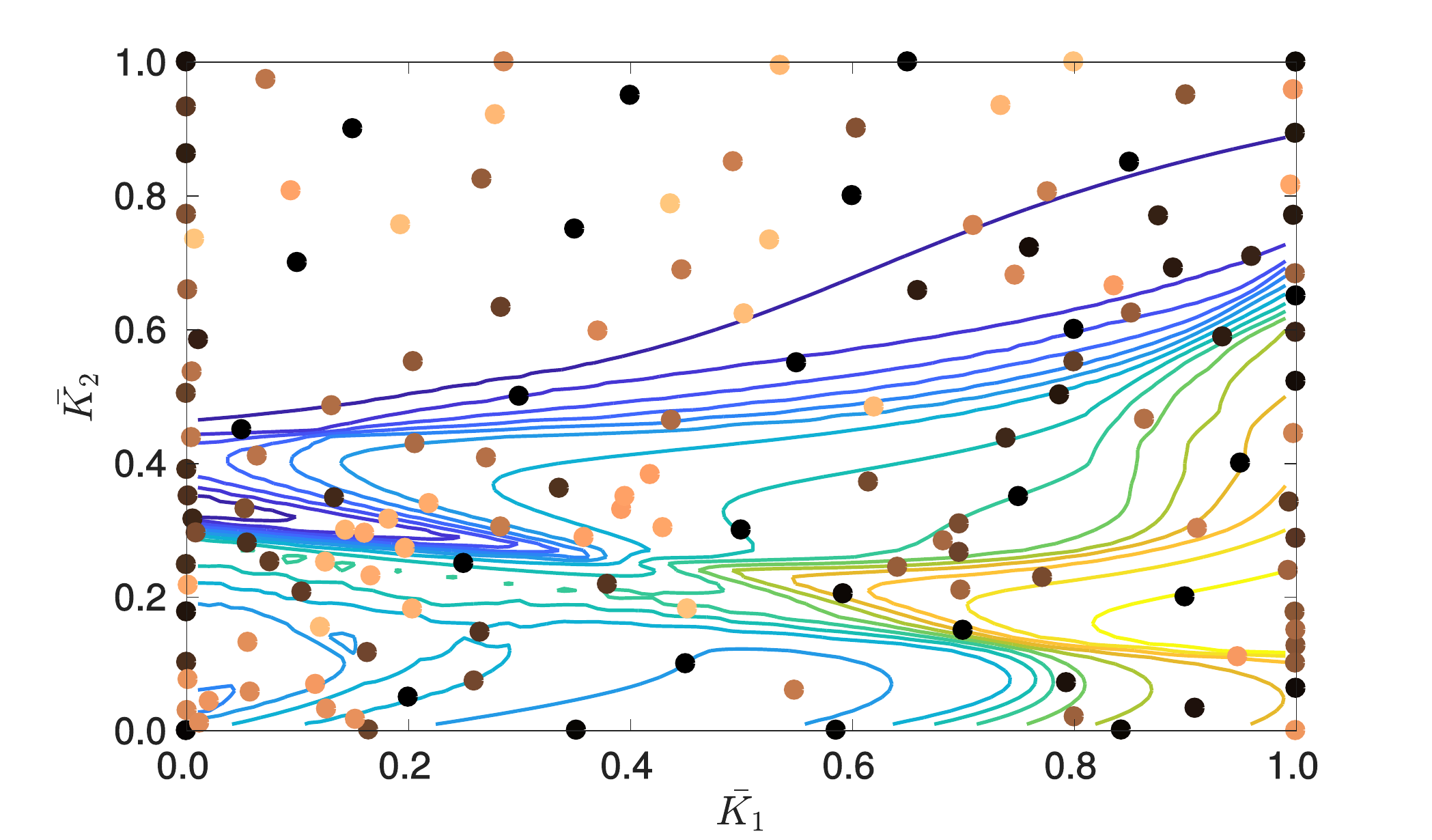}}

\caption[Sample positions and average number of samples for sticking sticking time problem]{Sample positions after 150 samples for the sticking time problem (Problem $\mathcal{P}_{0}$).}\label{fig:TStickKSeta06Yes}
\end{figure}

A set of sample positions of the added points of the two adaptive methods after 150 samples are shown in Figure \ref{fig:TStickKSeta06Yes}. On these plots the parameters have been normalized to lie between 0 and 1. The initial TPLHD sample points are shown as black circles. The later a point is added the more its color tends towards light red. The contour of the target function is displayed for both cases, it may be observed in Figure \ref{fig:TStickKSeta06YesMEPE} that MEPE balances proficiently exploration and exploitation, wheras EIGF focuses on specific domains.


\subsubsection{Comparison of kriging and other common metamodel techniques}

In this subsection ordinary kriging is compared to three commonly found surrogate modeling techniques, i.e. support-vector machines, radial basis function network and neural networks. Here, the metamodels are here considered without adaptive scheme, and are all based on the same sample set provided by \gls{tplhd}. For this comparison the support-vector machine is employed with a radial basis kernel as well as an automatic choice of the scale value for the kernel function as defined in the MATLAB software, see e.g. \cite{clarke2005analysis}. For the radial basis function network \citep{park1991universal}, the hidden layer has 50 neurons. The centers are calculated with k-means clustering algorithm and the widths are set to unity. The neural network is employed in form of multilayer perceptron based on 2 hidden-layer with 20 neurons each and sigmoid activation function where Levenberg-Marquardt backpropagation is used for training. 

The mean absolute errors as defined by Equation (\ref{eq:MAE}) for the obtained metamodels considering different sample sizes are listed in Table \ref{table::CompareMAE}. It can be noticed that the ranking of the methods depends on the number of samples considered. Support-vector machine performs better than radial basis function network and neural network for a low number of samples whereas it performs worse for number of samples larger than 100. Considering the parameters as detailed before, the results indicate that ordinary kriging yields the best approximation of the target function for all sample sizes  for the problem of interest, while simultaneously provding an estimate for the variance of the metamodel over the parametric space.
\begin{table}[ht!]
\begin{center}
\begin{tabular}{l c c c c c}
\hline
& \multicolumn{5}{c}{Number of samples} \\ 
\hline \\
Metamodel method  & 20 & 50 & 100 & 150 & 200  \\ \hline\hline \\
\bf{Ordinary kriging} &  \textbf{0.31} & \textbf{0.18} & \textbf{0.15} & \textbf{0.10} & \textbf{0.08}  \\
Support vector machine  &  0.37 & 0.32 & 0.28 & 0.26 & 0.25 \\
Radial basis function network & 0.46 & 0.46 &0.42 & 0.25 & 0.23  \\
Neural network & 0.84 & 0.47 & 0.32 & 0.23 & 0.16    \\ 
\end{tabular}
\end{center}
\caption[Comparison of the mean absolute error of different metamodeling techniques for the two dimensional sticking time]{Mean absolute errors in s for the two-dimensional sticking time problem (Problem $\mathcal{P}_{0}$) with four alternative metamodeling techniques and different sample sizes.}\label{table::CompareMAE}
\end{table}

\subsection{Surrogate model for classification of chaotic motion using \gls{lle}}\label{sec::SurModelForLLE}

The second focus of interest is to predict the parametric subdomains corresponding with chaotic motion of the oscillator at low computational cost using the kriging approach. It has been shown in \cite{Jan_Master_thesis,Class_Krig_sub} that issues arise when trying to utilize the common adaptive sampling techniques for a regressive metamodel for that application. Indeed, as previously discussed, the precise value of the \gls{lle} is not of particular interest. The main focus of a metamodel for this application is to classify $\hat{\mathcal{M}}_{LLE,MoB}(\bm{x})$ the approximated estimation of $\mathcal{M}_{LLE,MoB}(\bm{x})$ for a point $\bm{x}$ of the parametric domain either above or below 0 to characterize the behavior of the given dynamic system as chaotic or regular motion. Using the adaptive sampling technique MiVor recently proposed in \cite{Class_Krig_sub}, specifically dedicated to classification based on regression metamodel, is hereafter investigated to classify the \gls{lle} values. The required MiVor parameters (see \cite{Class_Krig_sub}), which are the initial exploration rate and the decrease factor, are chosen to be $0.4$ and $1.1$ respectively for all the following applications.

\subsection{One-dimensional classification for stick-slip instability: Problem $\mathcal{P}_{1}$}

\begin{figure}[htbp!]
\centering
\subfloat[\gls{lle} value and metamodel for classification]{\label{fig::LLE1d_plot}
\includegraphics[width=0.49\textwidth]{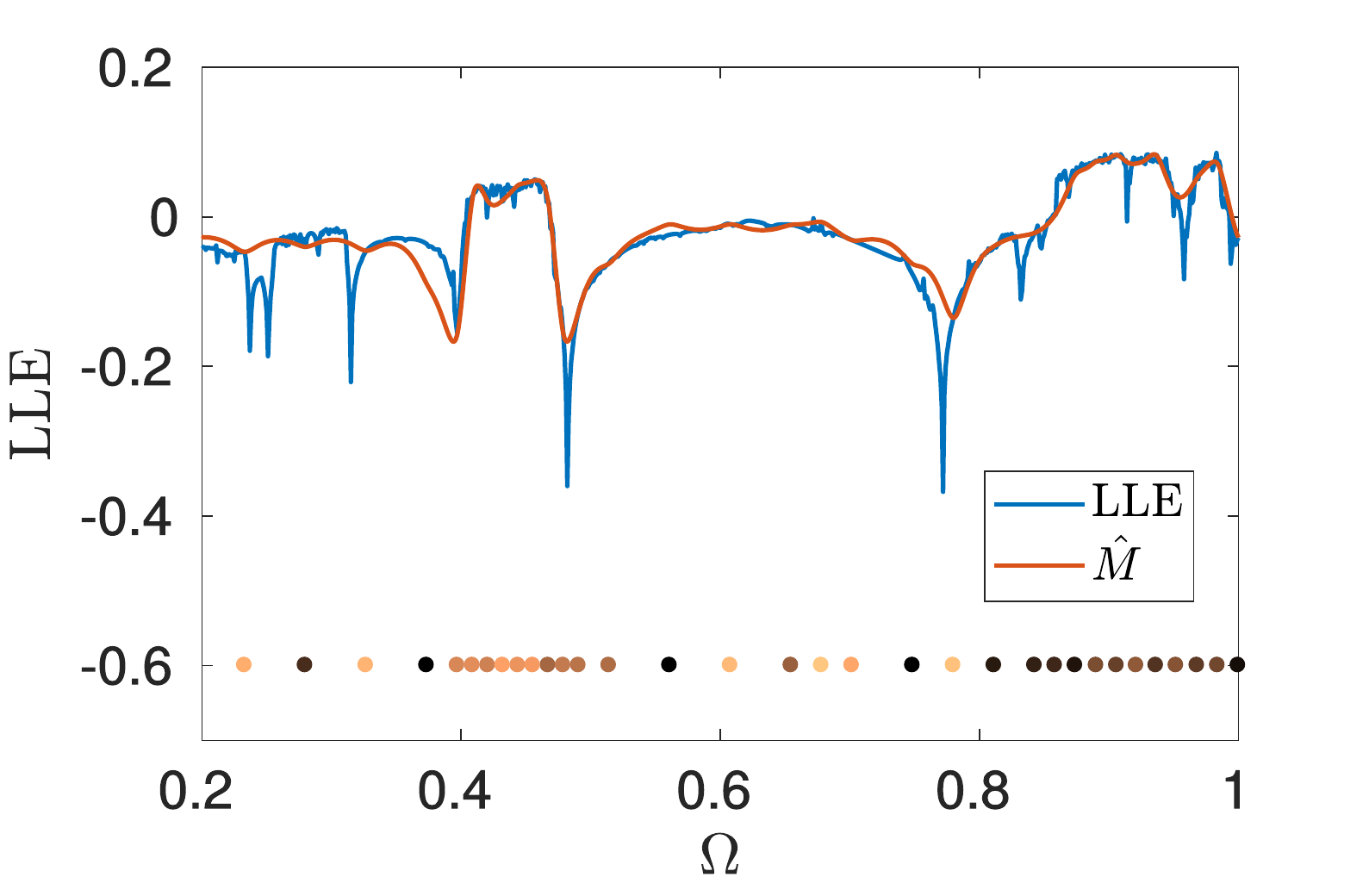}}
\subfloat[\footnotesize{Evolution of \gls{mivor} performance with number of samples}]{\label{fig:LLE1d_percent}
\includegraphics[width=0.49\textwidth]{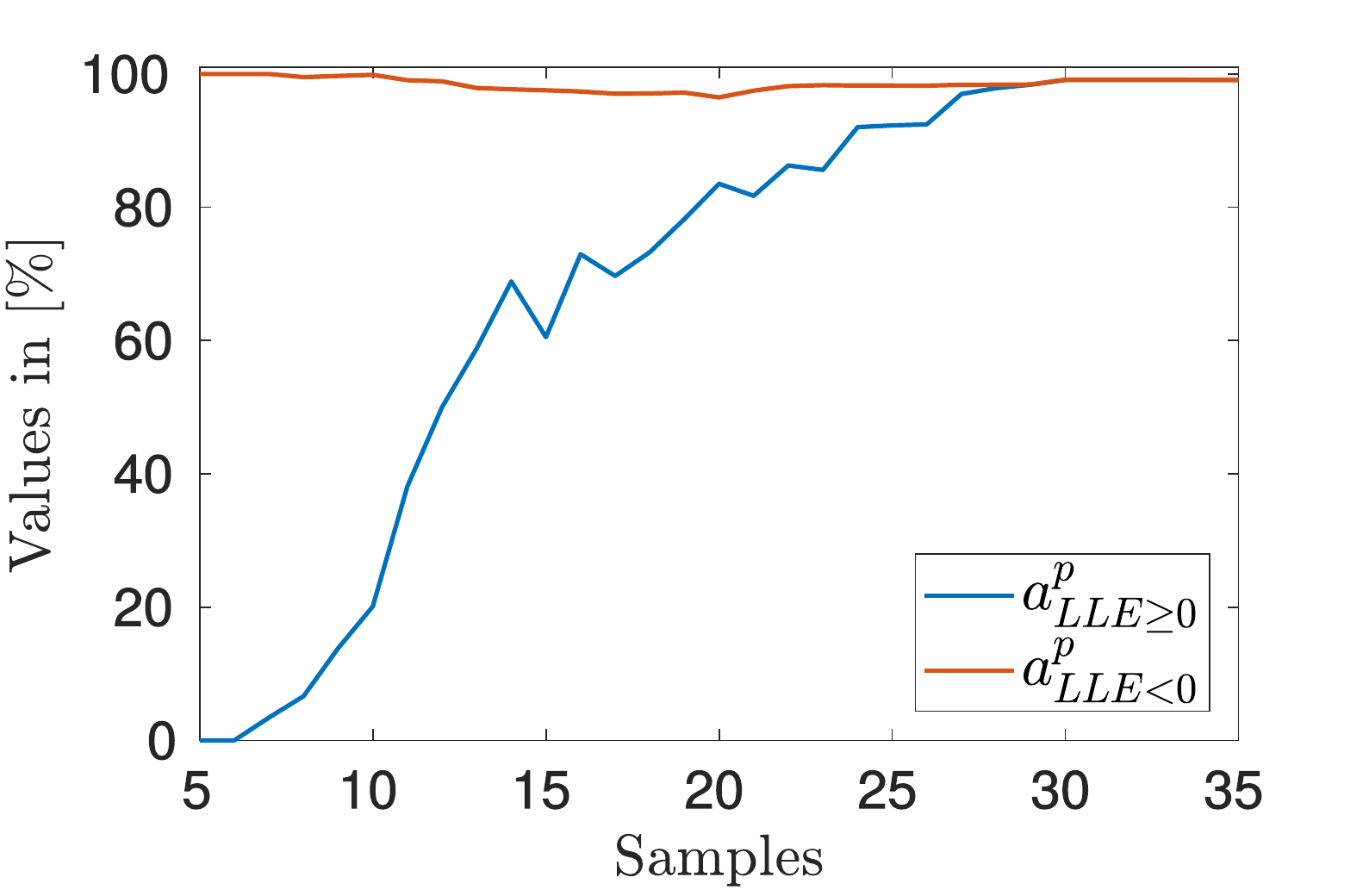}}
\caption{One-dimensional \gls{lle} example (Problem $\mathcal{P}_{1}$). $\Omega$ is given in $\text{rad}.\text{s}^{-1}$}
\label{fig::LLE1d}
\end{figure}

The first case of interest is the one-dimensional problem $\mathcal{P}_{1}$, where the goal of the metamodel is to identify the unstable behavior corresponding with the \gls{lle} values plotted in Figure \ref{fig::LLE1d_plot}. The angular frequency $\Omega$ varies between $0.2$ and $1.0\, \text{rad}.\text{s}^{-1}$, whereas the other input parameters have fixed values as defined in Table \ref{tab:parameters}.

 The performance of the metamodel is evaluated with respect to a reference solution obtained with 5000 sample points. The metamodel for classification is judged by how many of the points yielding $\mathcal{C}_{MoB}=0$ on one hand and how many points yielding $\mathcal{C}_{MoB}=1$ on the other hand are predicted. The measure is stated in percent of accurately predicted points. The percentage of correctly predicted points in the unstable regime is denoted by $ a^{p}_{\text{LLE} \, \geq \, 0} $ and defined as
 \begin{equation}
   a^{p}_{\text{LLE} \, \geq \, 0}  = \dfrac{\hat{n}^{\text{LLE} \, \geq \, 0}_{ref,\text{LLE} \, \geq \, 0}}{\text{n}_{ref,\text{LLE} \, \geq \, 0}}
 \end{equation}
 with $\hat{n}^{\text{LLE} \, \geq \, 0}_{ref,\text{LLE} \, \geq \, 0}$ the number of points among the ones predicted with unstable regime by the reference solution which are also predicted with unstable regime using the metamodel, and $\text{n}_{ref,\text{LLE} \, \geq \, 0}$ the number of points estimated in the unstable regime by the reference response surface. Furthermore, the percentage of correctly predicted points in the stable regime $ a^{p}_{\text{LLE} \, < \, 0}$ is given by
 \begin{equation}
   a^{p}_{\text{LLE} \, < \, 0}  = \dfrac{\hat{n}^{\text{LLE} \, < \, 0}_{ref,\text{LLE} \,<\, 0}}{\text{n}_{ref,\text{LLE} \,<\, 0}},
 \end{equation}
 where $\text{n}_{ref,\text{LLE} \,<\, 0}$ is the number of points with regular behavior given by the reference response surface. Besides $\hat{n}^{\text{LLE} \, < \, 0}_{ref,\text{LLE} \,<\, 0}$ denotes the number of points estimated with regular behavior using the kriging approach among the points which are predicted with regular behavior using the reference response surface. 
 
 The metamodel is built from the prior information given by five initial sample points set with \gls{tplhd}, i.e. $[0.2,0.36,0.52, 0.68, 0.84]^{T}$. None of them yields a chaotic output. The challenged posed to the adaptive metamodeling strategy is to evaluate if the scheme is able to explore the parametric domain to identify the chaotic subdomain.
 
 The reference solution, the metamodel $\hat{M}$ after 35 samples and the respective sample positions at the end of the adaptive process are shown in Figure \ref{fig::LLE1d_plot}. It can be seen that a proficient metamodel has been created with only few observation points. Furthermore as intended with \gls{mivor} most of the created samples lie in the area that indicates chaotic behavior. 
 
 The evolution of the accuracy with regard to the sample size is shown in Figure \ref{fig:LLE1d_percent}. As an arbitrarily chosen stopping criterion the procedure stops after 35 samples, i.e. after adding 30 observation points to the initial samples. Finally, the metamodel correctly evaluates $99.11\,\%$ of the 5000 random points that yield LLE $\geq 0$ and $99.23\,\%$ of the points with LLE $<0$.

\subsubsection{Problem $\mathcal{P}_{2}$}

Next, the metamodel approach is tested for the test problem $\mathcal{P}_{2}$ which aims at evaluating the ability of the proposed approach to proficiently explore the parametric domain if only a small percentage of the domain has a chaotic motion.

\begin{figure}[htbp!]
\centering
\subfloat[$\mathcal{M}_{LLE,MoB} $]{\label{fig:KSETA06SmallPlot}
\includegraphics[width=0.49\textwidth]{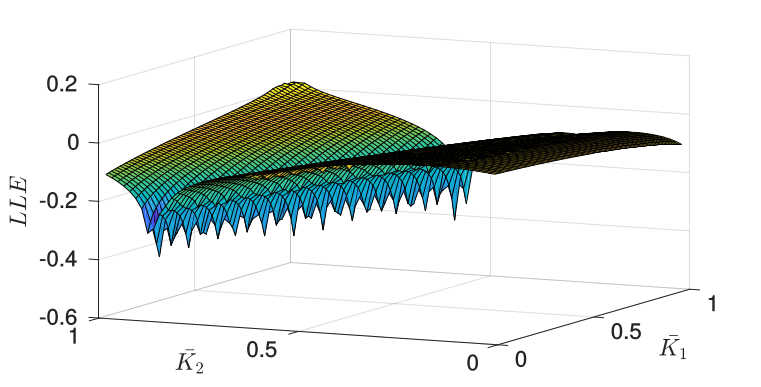}}
\subfloat[Classification provided by $\mathcal{M}_{LLE,MoB} $]{\label{fig:KSETA06SmallRedGray}
\includegraphics[width=0.49\textwidth]{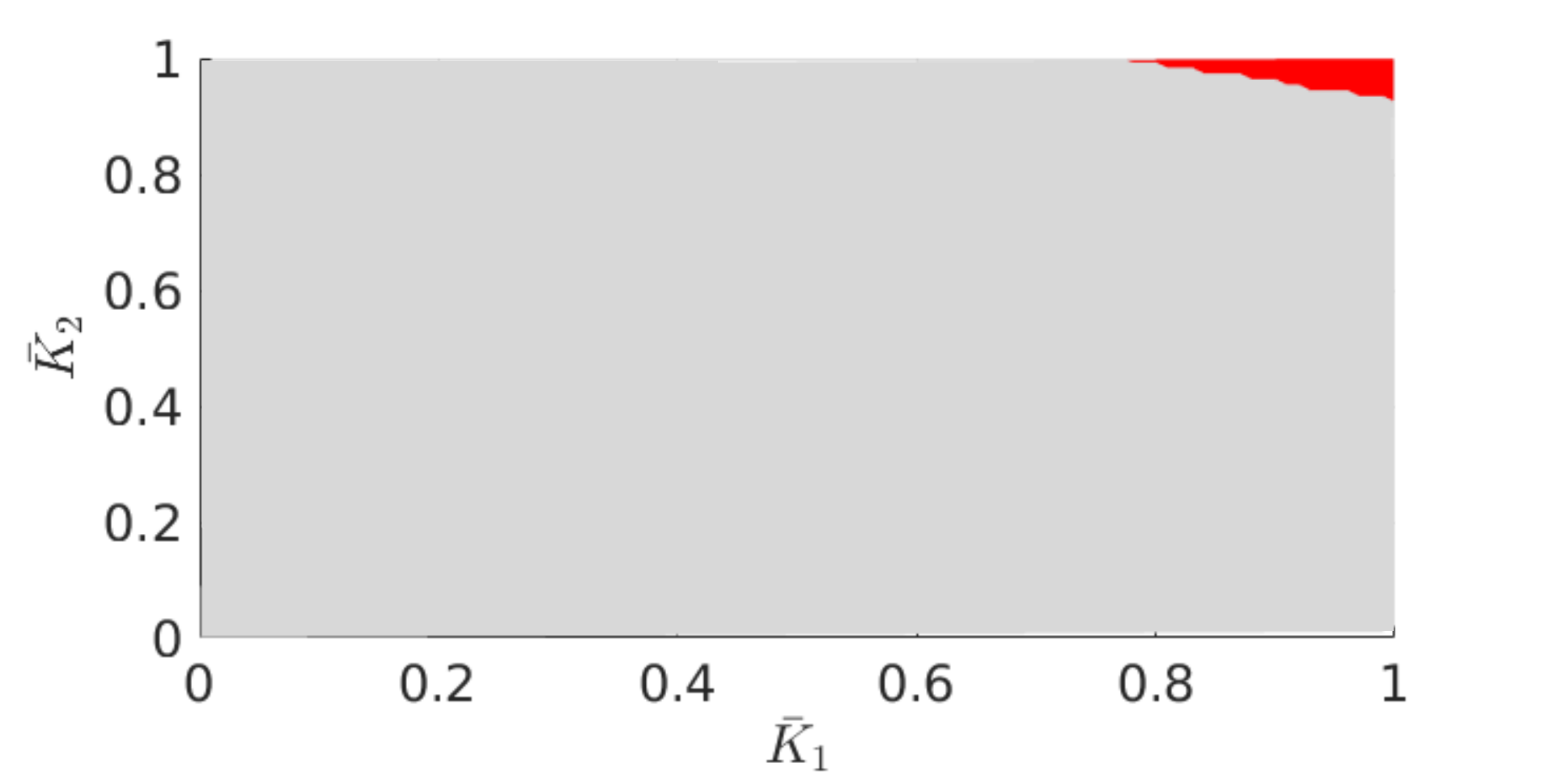}}

\subfloat[65 \gls{mivor} samples]{\label{fig:KSETA06SmallSamples}
\includegraphics[width=0.49\textwidth]{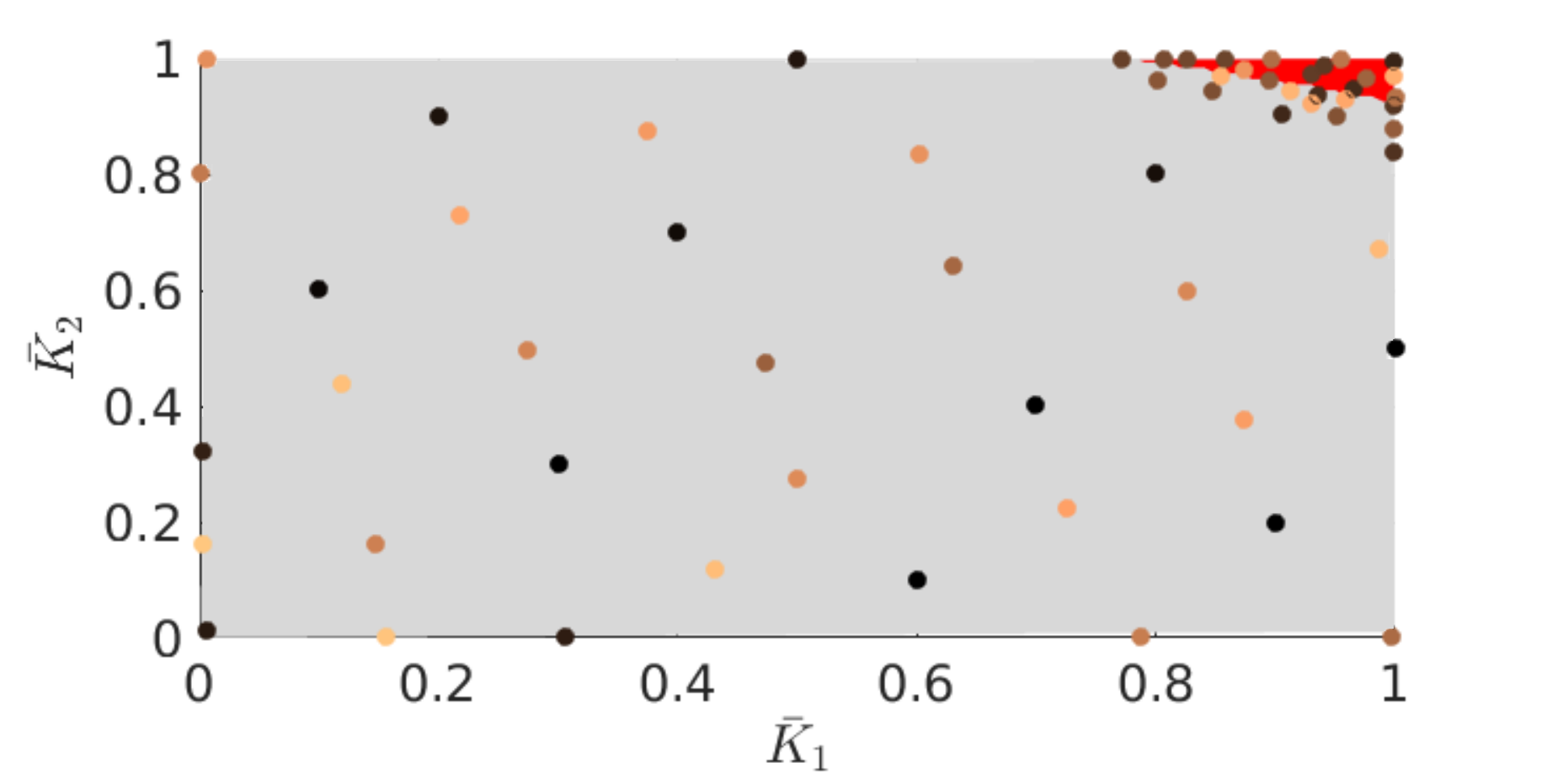}}
\subfloat[Classification provided by $\hat{\mathcal{M}}_{LLE,MoB} $ with 60 samples]{\label{fig:KSETA06SmallMeta}
\includegraphics[width=0.49\textwidth]{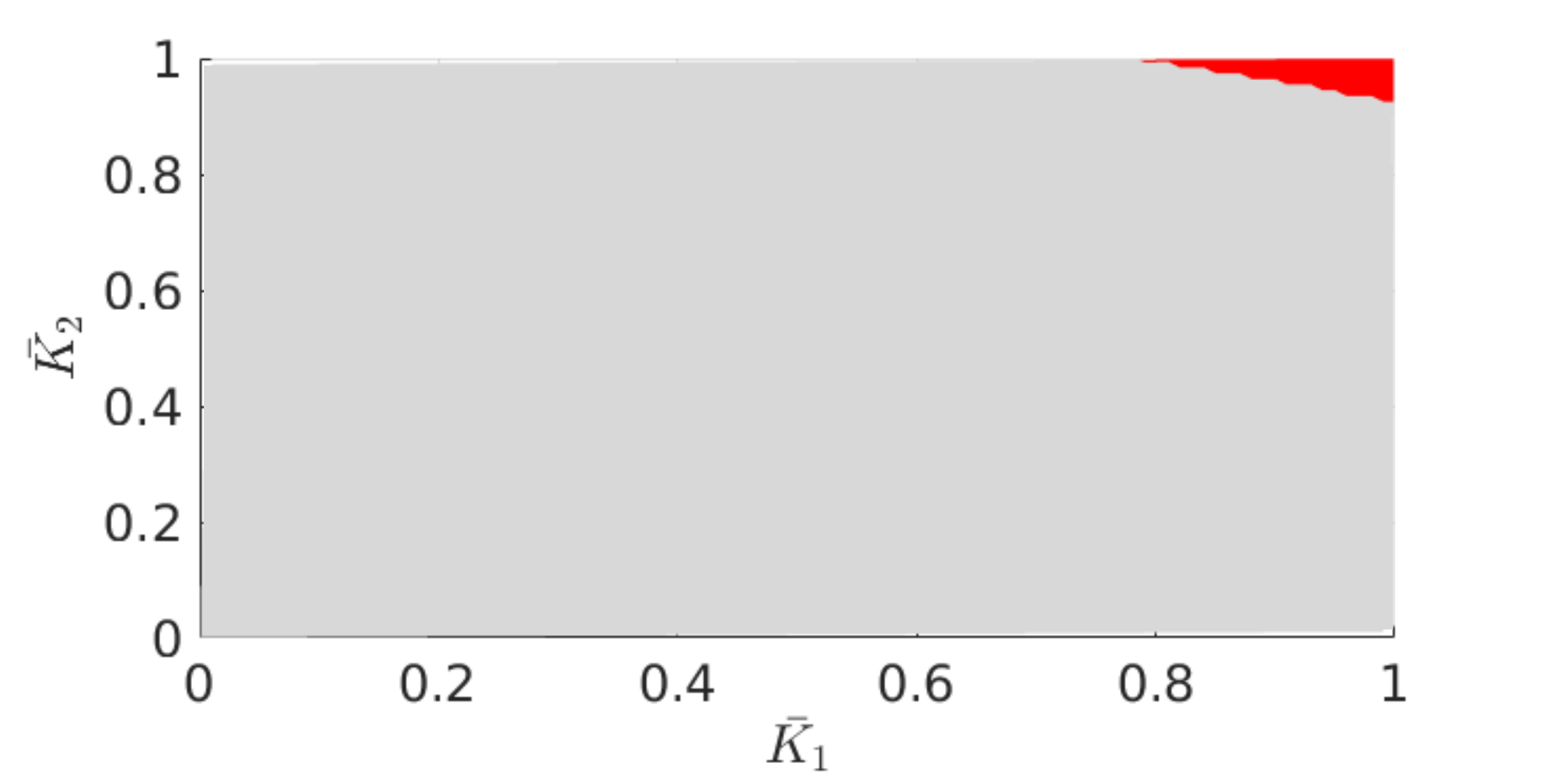}}
\caption[Results and data for 2D \gls{lle} case $\mathcal{P}_{2}$]{LLE response surface, LLE classes and metamodel approximation and observations for the 2D LLE case $\mathcal{P}_{2}$. $\bar{K}_{1}$ and $\bar{K}_{2}$ are the normalized spring stiffnesses.}\label{fig:KSETA06Small}
\end{figure}

Consider the two-dimensional input domain for the two spring stiffnesses given by $K_{1} \in  [0.5 , 1.0]\, \text{N}.\text{m}^{-3}$
and $K_{2} \in [0.0,0.5] \, \text{N}.\text{m}^{-1}$, see Table \ref{tab:parameters} for the values of other parameters.
The plot of the reference response surface $\mathcal{M}_{LLE,MoB}$ provided by 10000 computations which were spread over the input domain in a space-filling manner using TPLHD is shown in Figure \ref{fig:KSETA06SmallPlot}. Large fluctuations on the LLE values can be observed. The analysis of the parametric domain through binary classification corresponding with either stable or chaotic motion is plotted in Figure \ref{fig:KSETA06SmallRedGray}. As before, regular motion is represented in grey whereas the red area informs about chaotic behavior. It can be seen that only a small fraction of the domain yields chaotic motion. Among the 10000 points computed to build the reference solution only 93 points lead to stick-slip instability, i.e. only less than $1 \, \%$ of the points. 

The metamodel is built from an initial set of 10 initial experiments created with \gls{tplhd}, then adaptive sampling is used until reaching 65 samples. The point locations of the 55 samples created with \gls{mivor} over the normalized input space are highlighted in Figure \ref{fig:KSETA06SmallSamples}. It can be observed that they are predominantly spread in and around the red area that indicates chaos. This leads to an accurate surrogate model $\hat{\mathcal{M}}_{LLE,MoB} $ for the classification problem as shown in Figure \ref{fig:KSETA06SmallMeta}. In details, it yields that $98.93 \, \% $ of the points above or equal 0 are correctly classified and $99.91 \, \%$ of the points below 0.

\begin{figure}[htbp!]
\centering
\includegraphics[width=0.7\textwidth]{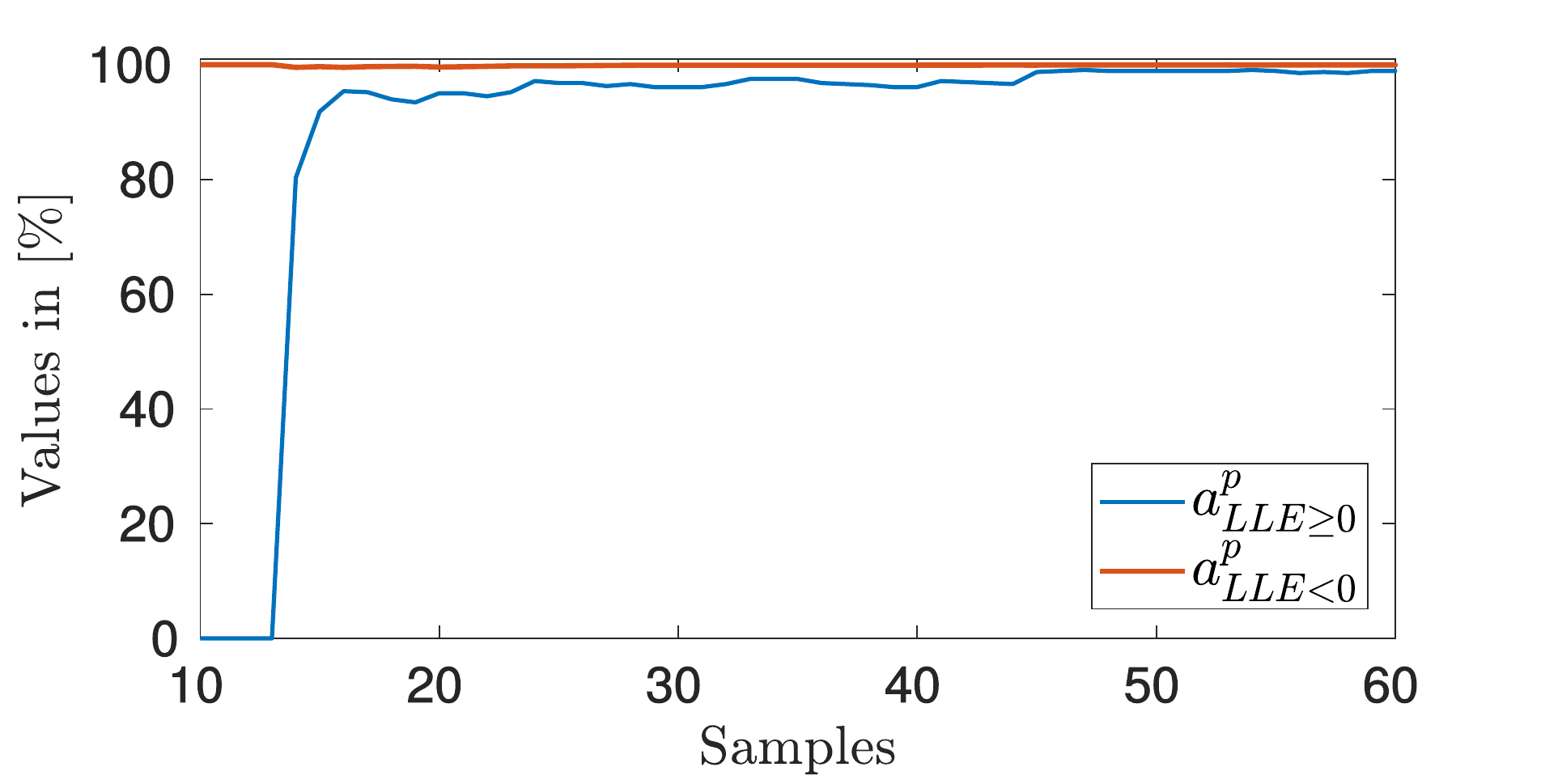}
\caption[Convergence of correctly classified points for 2D LLE case $\mathcal{P}_{2}$]{Convergence of correctly classified points for 2D LLE case $\mathcal{P}_{2}$.}\label{fig::KSeta06SmallConvergence}
\end{figure}

The convergence of the error measures for \gls{mivor} is depicted in Figure \ref{fig::KSeta06SmallConvergence}. It can be seen that $a^{p}_{LLE \geq 0}$ \gls{mivor} shows a big jump around 15 sample points. This is due to the fact that the algorithm needs to identify the small area of Figure \ref{fig:KSETA06SmallRedGray}. After the jump the metric quickly converges to the optimal state.  As a large part of the \gls{lle} domain is below the threshold value $a^{p}_{LLE < 0}$ appears almost accurate from the beginning with values varying between $99.6-100\,\%$.

\subsubsection{Problem $\mathcal{P}_{3}$}

Problem $\mathcal{P}_{3}$ aims at evaluating metamodel performance to detect almost disconnected regions of the input domain associated with chaotic behavior. 

\begin{figure}[htbp!]
\centering
\subfloat[$\mathcal{M}_{LLE,MoB} $]{\label{fig:muKK2Plot}
\includegraphics[width=0.49\textwidth]{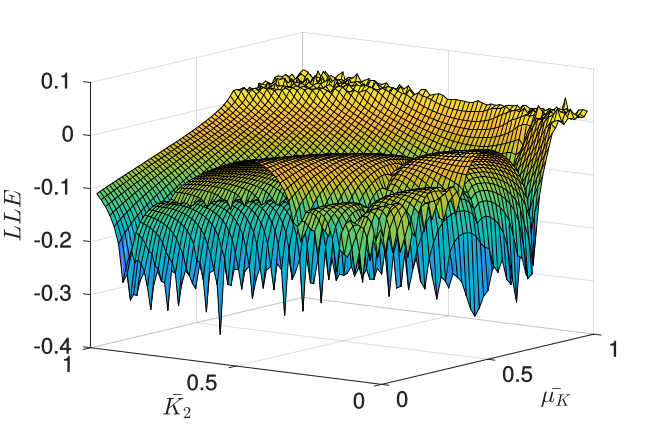}}
\subfloat[Classification provided by $\mathcal{M}_{LLE,MoB} $]{\label{fig:muKK2RedGray}
\includegraphics[width=0.49\textwidth]{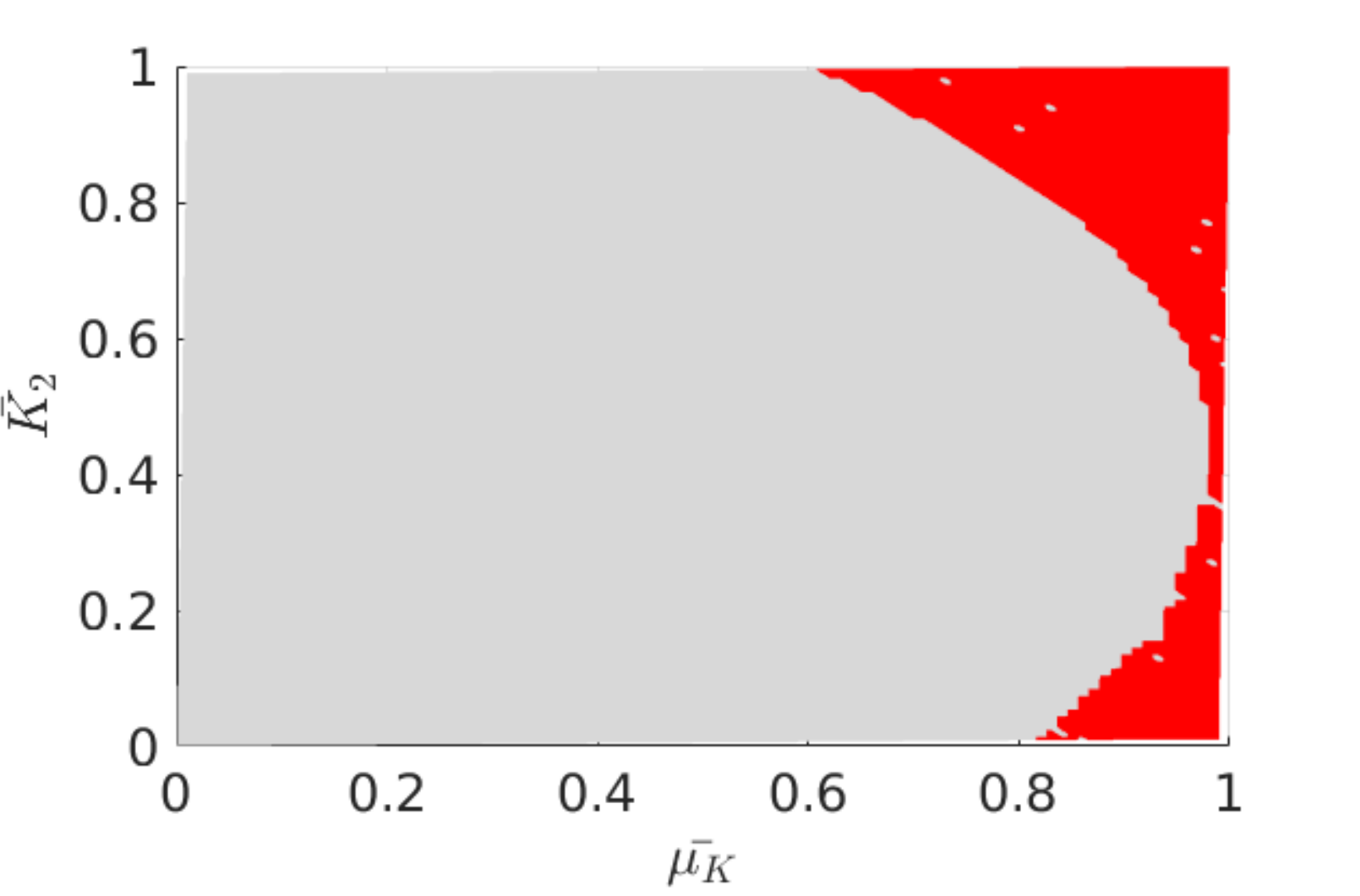}}

\subfloat[105 \gls{mivor} samples]{\label{fig:muKK2Samples}
\includegraphics[width=0.49\textwidth]{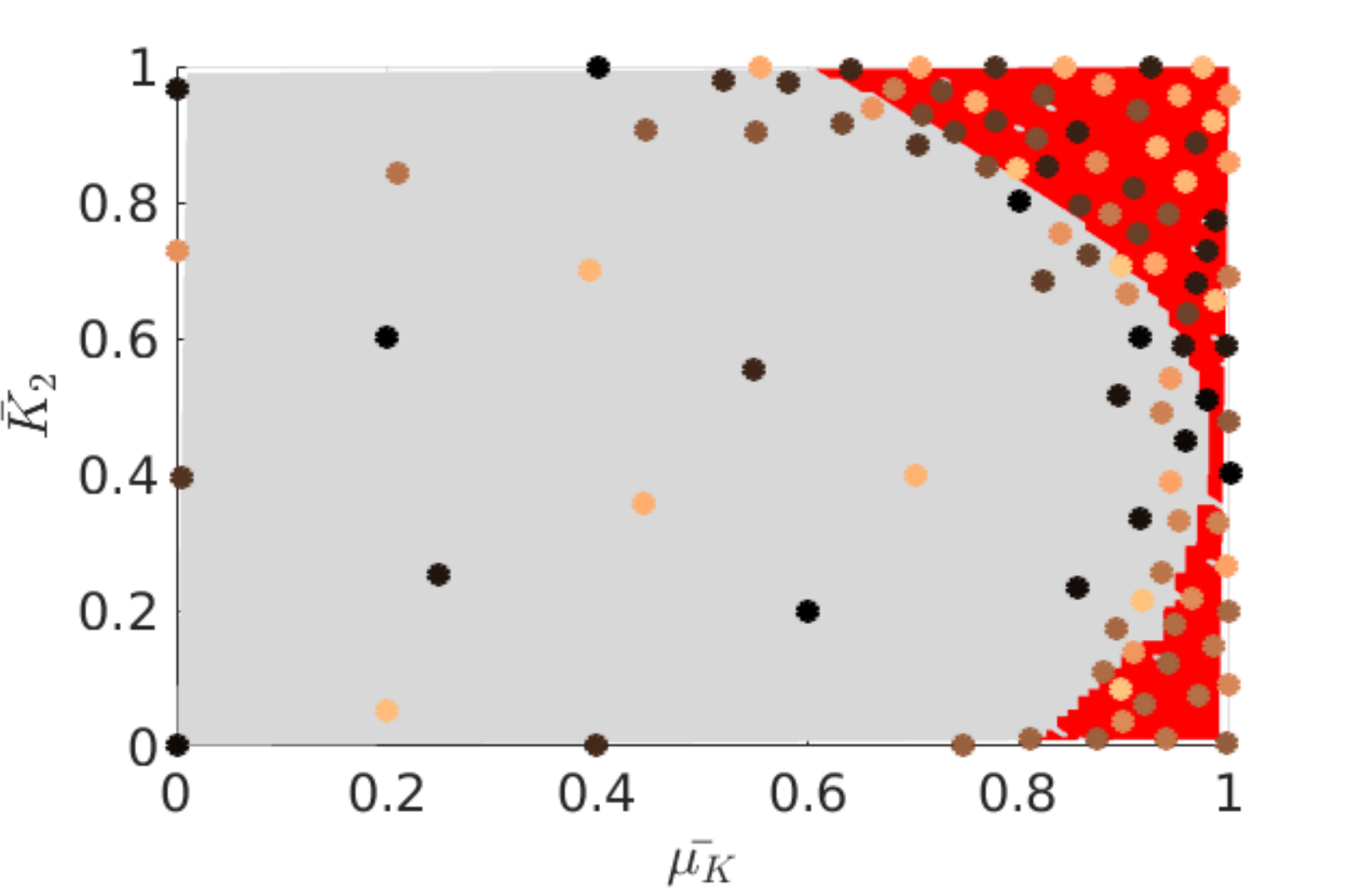}}
\subfloat[Classification provided by $\hat{\mathcal{M}}_{LLE,MoB} $ with 105 samples]{\label{fig:muKK2Meta}
\includegraphics[width=0.49\textwidth]{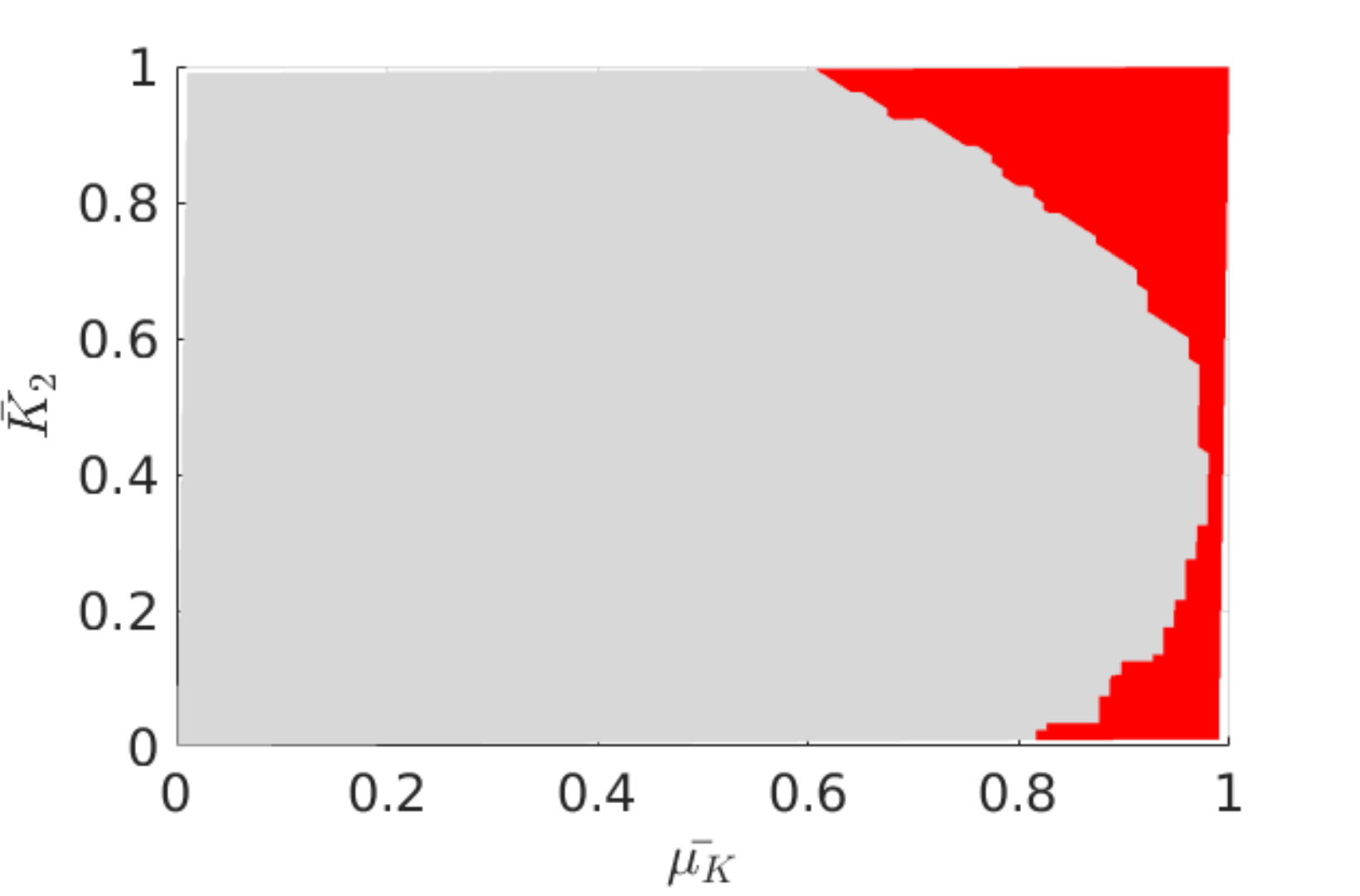}}
\caption[Results and data for 2D LLE case $\mathcal{P}_{3}$]{LLE response surface, LLE classes, metamodel approximation and observations for the 2D LLE case $\mathcal{P}_{3}$. }\label{fig:muKK2}
\end{figure}

Consider the input domain for the spring stiffness given by \\
$K_{2} \in [0.0,0.5] \, \text{N}.\text{m}^{-1}$ and the kinematic friction coefficient $\mu_{k} \in [0.08,0.18]$. Values of all other parameters are listed in Table \ref{tab:parameters}. The LLE reference surface established from 10000 sample points is plotted in Figure \ref{fig:muKK2Plot}. It exhibits large fluctuations. This reference solution yields the classification displayed in Figure \ref{fig:muKK2RedGray}. It can be seen that there are two chaotic subdomains which are almost disconnected. 

\begin{figure}[htbp!]
\centering
\includegraphics[width=0.7\textwidth]{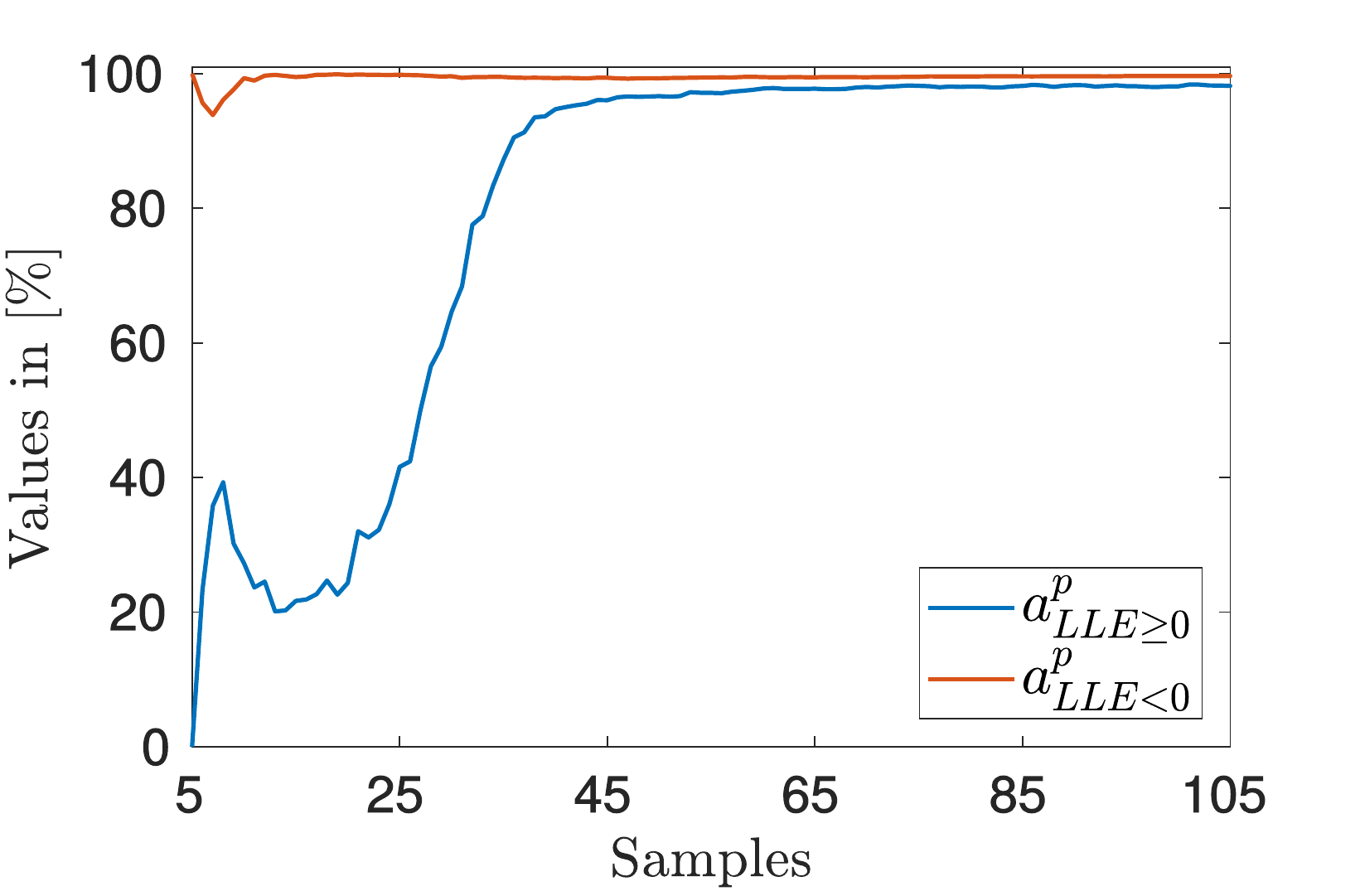}
\caption[Convergence of correctly classified points with for 2D LLE problem $\mathcal{P}_{3}$]{Convergence of correctly classified points for 2D LLE problem $\mathcal{P}_{3}$.}\label{fig::P3_convergence}
\end{figure}

The metamodel is constructed adaptively from 5 points sampled with TPLHD. It is evaluated after adding 100 sample points and the results are averaged over 20 independent realizations. The locations of the 105 \gls{mivor} sample points for one realization are shown in Figure \ref{fig:muKK2Samples}. It can be noticed that the method is effective in sampling around the red zones indicating chaos. Furthermore the rest of the domain is sampled with a space-filling design. The resulting metamodel after 105 samples is displayed in Figure \ref{fig:muKK2Meta} and shows an accurate prediction performance. The percentage of correctly identified points for LLE$\geq 0$ is $98.53\%$, whereas $99.19\%$ of the points corresponding to LLE values below 0 are correctly identified. The convergence of the percentage values is displayed in Figure \ref{fig::P3_convergence}. It shows a very proficient sampling performance of \gls{mivor}.

\subsubsection{Problem $\mathcal{P}_{4}$}

The metamodel approach for identifying stick-slip instability is tested on Problem $\mathcal{P}_{4}$, characterized by a highly complex response surface, in which regular and chaotic motion subdomains are embedded into each other. Consider the parametric domain for the two spring stiffnesses given by $K_{1} \in  [0.5 , 1.0]\, \text{N}.\text{m}^{-3}$
and $K_{2} \in [0.0,0.6] \, \text{N}.\text{m}^{-1}$ with other parameters as given in Table \ref{tab:parameters}.

\begin{figure}[htbp!]
\centering
\subfloat[$\mathcal{M}_{LLE,MoB}$]{\label{fig::Kseta07Plot}
\includegraphics[width=0.49\textwidth]{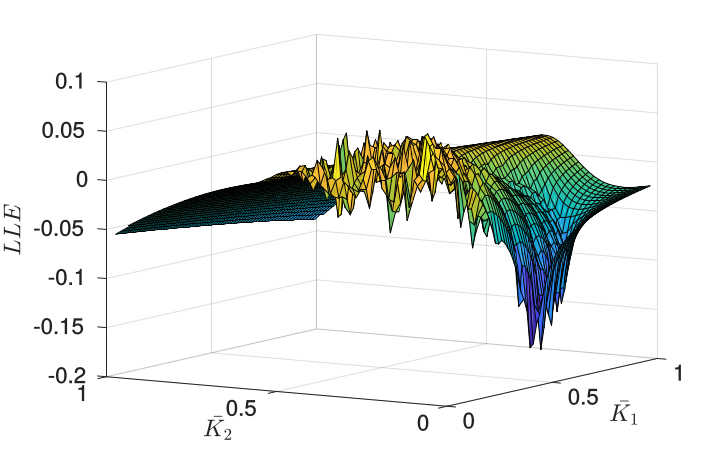}}
\subfloat[Classification provided by $\mathcal{M}_{LLE,MoB} $]{\label{fig::Kseta07redGray}
\includegraphics[width =0.49\textwidth]{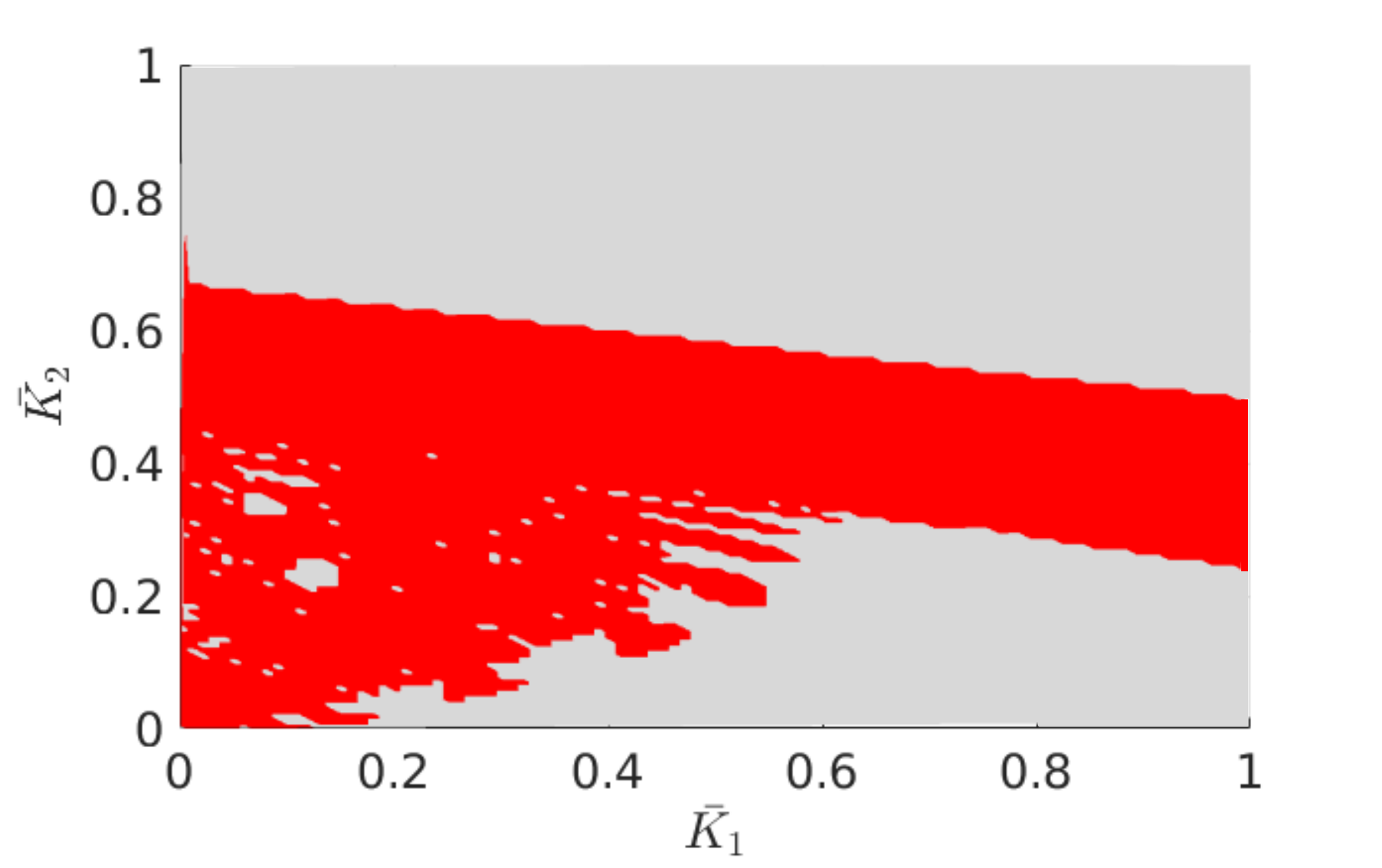}}

\subfloat[115 \gls{mivor} samples]{\label{fig::Kseta07Samples}
\includegraphics[width =0.49\textwidth]{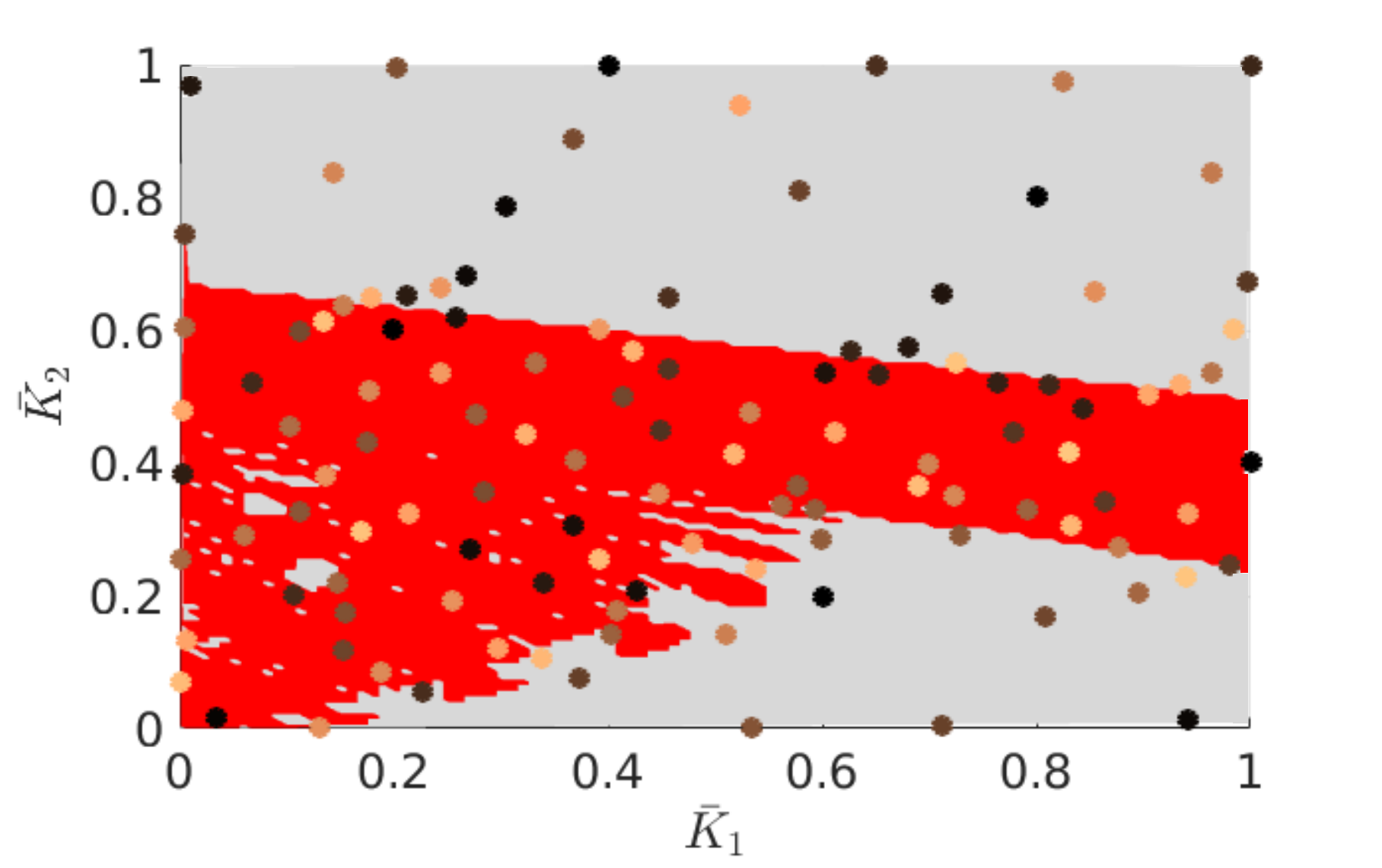}}
\subfloat[Classification provided by $\hat{\mathcal{M}}_{LLE,MoB} $ with 115 samples]{\label{fig::Kseta07Meta}
\includegraphics[width =0.49\textwidth]{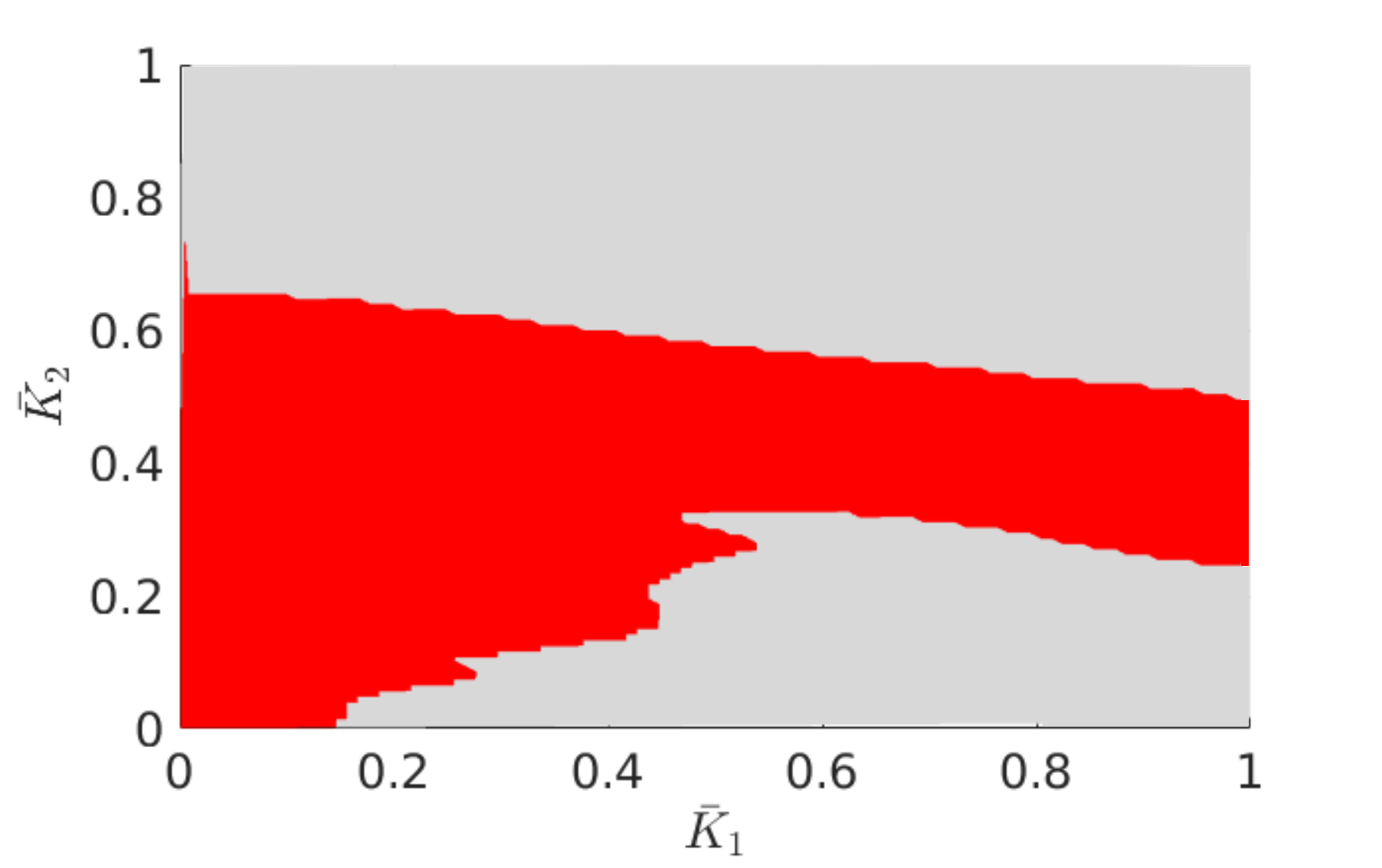}}
\caption[Results and data for 2D LLE case $\mathcal{P}_{4}$]{LLE response surface, LLE classes, metamodel approximation and observations for the 2D LLE case $\mathcal{P}_{4}$. $\bar{K}_{1}$ and $\bar{K}_{2}$ are the normalized spring stiffnesses. LLE is unitless.}\label{fig::Kseta07}
\end{figure}

The reference LLE response surface based on the evaluation of 10000 sample points is shown in Figure \ref{fig::Kseta07Plot}. The corresponding classification is pictured in Figure \ref{fig::Kseta07redGray}. It can be noticed that the chaotic area is complex with some regular subdomains embedded within chaotic behavior.

A set of 5 points sampled using the TPLHD technique is considered as initial observation points. One realization is represented in Figures \ref{fig::Kseta07Samples} and \ref{fig::Kseta07Meta}. The adaptive sampling technique is evaluated after reaching 115 sample points. Samples needed to generate the metamodel are shown in Figure \ref{fig::Kseta07Samples}. The resulting metamodel is represented in Figure \ref{fig::Kseta07Meta}. It can be seen that the holes inside the red domain are not detected by the surrogate model. This reduces the percentage of correctly classified points. However an accurate prediction of these holes with kriging would require a significantly larger number of samples. Furthermore identifying a point inside the regular motion subdomain might lead to a perceived reliability of the system dynamics around the observation point which might not be desired from an engineering point of view. 

\begin{figure}[htbp!]
\centering
\includegraphics[width=0.7\textwidth]{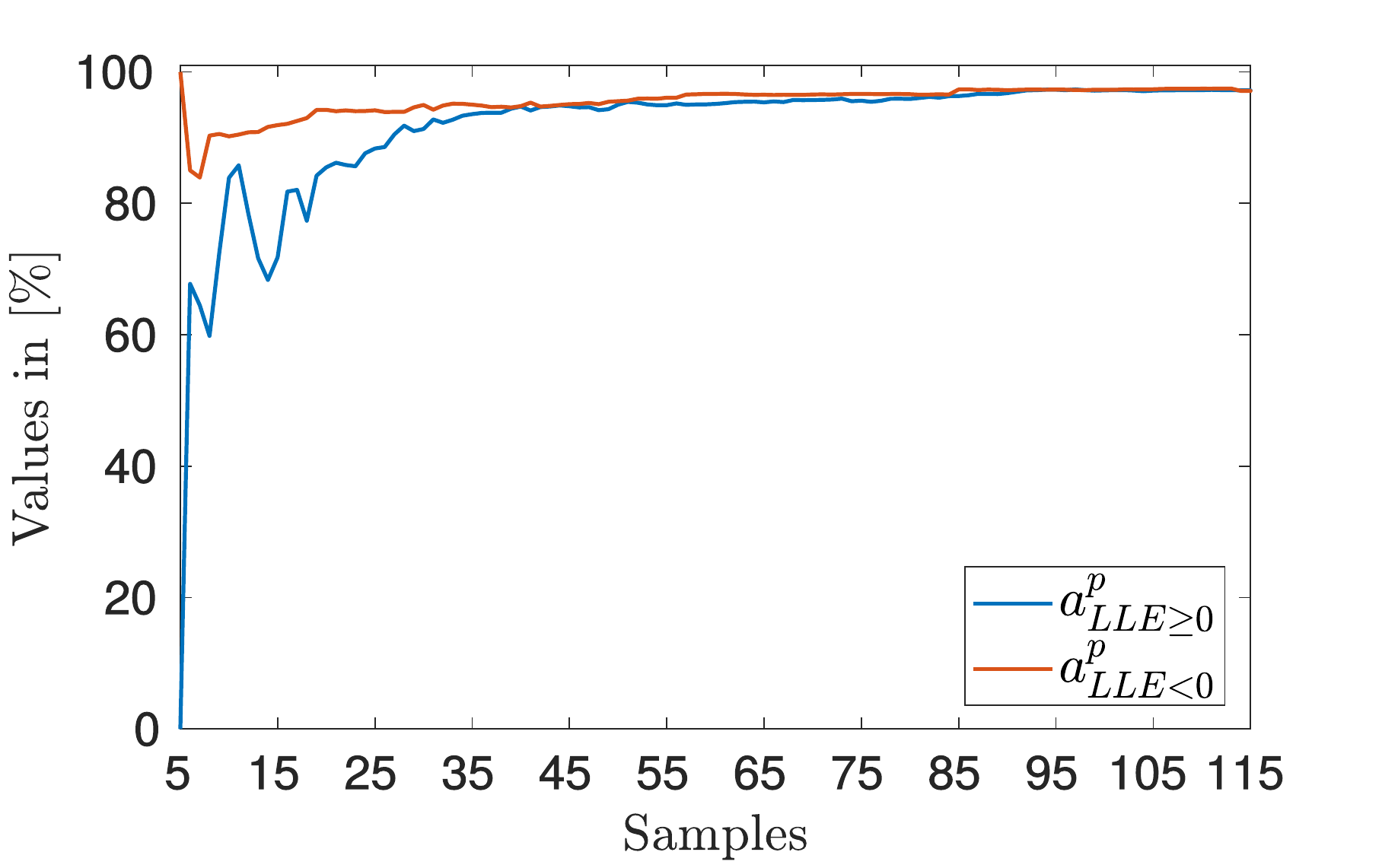}
\caption[Convergence of correctly classified points with for 2D LLE case $\mathcal{P}_{4}$]{Convergence of correctly classified points for 2D LLE problem $\mathcal{P}_{4}$.}\label{fig::P4_convergence}
\end{figure}

The convergence of the error metrics are displayed in Figure \ref{fig::P4_convergence}, which shows an efficient convergence behavior. The results of the \gls{mivor} metamodel after 115 samples are very proficient with values of $a^{p}_{LLE \geq 0}= 97.12 \, \%$ and $a^{p}_{LLE < 0} = 97.32 \, \%$. Considering the very complex form of the chaotic region including notched boundary and holes as shown in Figure \ref{fig::Kseta07redGray}, these percentages of accurately identified points computing only 115 observation points are highly promising for detecting non-stable dynamic parametric domains at low cost.

\subsection{Problem $\mathcal{P}_{5}$}

Finally the input dimension is extended to the three-dimensional problem $\mathcal{P}_{5}$. Consider the parametric space  $\Omega \times  K_{2} \times \mu_{K} \in [0.6, 0.9] \, \text{rad}.\text{s}^{-1} \times [0.0, 0.5] \, \text{N}.\text{m}^{-1}$ $\times [0.1, 0.15]  $ with $K_{1} = 1 \, \text{N}.\text{m}^{-3}$. The rest of the parameters remain unchanged with regard to the previous examples as given in Table \ref{tab:parameters}. As reference 15000 observations, which are evenly spread in the parametric space, are evaluated. 

\begin{figure}[h!]
\centering
\includegraphics[width=0.7\textwidth]{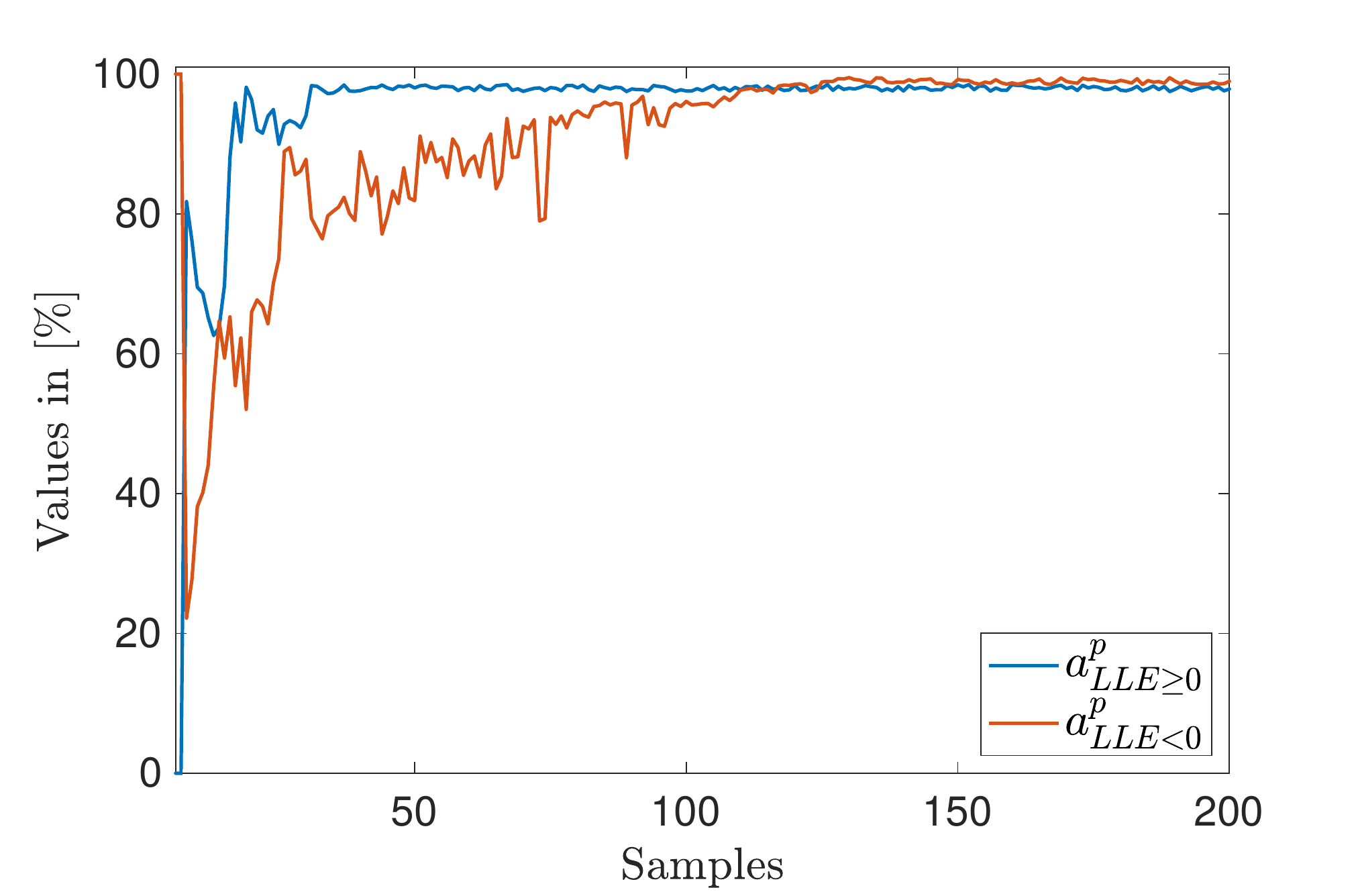}
\caption[Convergence of correctly classified points with for 3D LLE case $\mathcal{P}_{4}$]{Convergence of correctly classified points for 3D LLE problem $\mathcal{P}_{5}$.}\label{fig::P5_convergence}
\end{figure}

The \gls{mivor} process is started from an initial set of 5 TPLHD samples, none of which correspond to an unstable dynamic system.  The evolution of the averaged error values until including $200$ samples is displayed in Figure \ref{fig::P5_convergence}.

It can be noticed that \gls{mivor} achieves proficient value convergence with both metrics indicating an exactitude of above $98 \%$ of accurately identified points for both classes with already around 125 observation points. The jittering of the values indicates sharp drops of the LLE response surface.
It can be seen that \gls{mivor} can be very well expanded to higher dimension on that specific example.

\section{Conclusion}
Using the latest development proposed in the literature both in terms of numerical approaches allowing to compute the \gls{lle} and kriging models for classification, a proficient analysis of the parametric model can be conducted to optimally improve the knowledge of dynamic instabilities. A non-linear oscillator of Duffing's type with an elasto-plastic friction model has been investigated. The largest Lyapunov exponent as well as the sticking time have been introduced as means to investigate the system behavior. The analysis of this system which appeared highly challenging, as both QoIs have highly fluctuating values in the parametric domain, is very efficiently tackled using an adaptive metamodeling approach. The LLE appears as the most promising QoI in that context. Values equal or above zero indicate chaotic motion accurately. The \gls{mivor} sampling technique is able to generate surrogate models able to classify chaotic and regulate motion with only a few observations.  It efficiently gains information from the available observation to further guide the investigation and thus allows to identify the instability domain(s) with reasonable number of computations. It therefore provides pertinent information for design optimization, even for complex response surfaces and two- or three-dimensional parametric spaces.

\begin{acknowledgements}
The authors acknowledge the financial support from the Deutsche Forschungsgemeinschaft under
Germany’s Excellence Strategy within the Cluster of Excellence
PhoenixD (EXC 2122, Project ID 390833453).\\
The results presented in this paper were partially carried out on the cluster system at the Leibniz University of Hannover, Germany.
\end{acknowledgements}

\textbf{Conflict of interest} The authors declare that they have no conflict of interest.


\bibliographystyle{unsrt}   
\bibliography{bib}

\begin{thebibliography}{10}

\bibitem{rabinowicz1956stick}
E~Rabinowicz.
\newblock Stick and slip.
\newblock {\em Scientific American}, 194(5):109--119, 1956.

\bibitem{barton2004braking}
D~Barton and A~Blackwood.
\newblock {\em Braking 2004: Vehicle Braking and Chassis Control}, volume~6.
\newblock John Wiley \& Sons, 2004.

\bibitem{ashraf2017investigation}
N~Ashraf, D~Bryant, and {JD} Fieldhouse.
\newblock Investigation of stick-slip vibration in a commercial vehicle brake
  assembly.
\newblock {\em International Journal of Acoustics \& Vibration}, 22(3), 2017.

\bibitem{owen2003reduction}
William~S Owen and Elizabeth~A Croft.
\newblock The reduction of stick-slip friction in hydraulic actuators.
\newblock {\em IEEE/ASME transactions on mechatronics}, 8(3):362--371, 2003.

\bibitem{wu2017comparisons}
Q~Wu, S~Luo, Qu~T, and X~Yang.
\newblock Comparisons of draft gear damping mechanisms.
\newblock {\em Vehicle System Dynamics}, 55(4):501--516, 2017.

\bibitem{rubio2007structural}
D~Rubio and L~{San Andres}.
\newblock Structural stiffness, dry friction coefficient, and equivalent
  viscous damping in a bump-type foil gas bearing.
\newblock {\em Journal of engineering for gas turbines and power},
  129(2):494--502, 2007.

\bibitem{stelter1990stick}
P~Stelter.
\newblock Stick-slip vibrations and chaos.
\newblock {\em Phil. Trans. R. Soc. Lond. A}, 332(1624):89--105, 1990.

\bibitem{hinrichs1997dynamics}
N~Hinrichs, M~Oestreich, and K~Popp.
\newblock Dynamics of oscillators with impact and friction.
\newblock {\em Chaos, Solitons \& Fractals}, 8(4):535--558, 1997.

\bibitem{galvanetto1999dynamics}
U~Galvanetto and S~Bishop.
\newblock Dynamics of a simple damped oscillator undergoing stick-slip
  vibrations.
\newblock {\em Meccanica}, 34(5):337--347, 1999.

\bibitem{jimenez2007two}
M~Jim{\'e}nez, J~Bielsa, R~Rodr{\'\i}guez, and C~Bernad.
\newblock Two {FEM} approaches for the prediction and quantification of
  “stick-slip” phenomena on rubber-metal sliding contacts.
\newblock In {\em IUTAM Symposium on Computational Methods in Contact
  Mechanics}, pages 291--309. Springer, 2007.

\bibitem{devarajan2017analytical}
K~Devarajan and B~Balaram.
\newblock Analytical approximations for stick--slip amplitudes and frequency of
  duffing oscillator.
\newblock {\em Journal of Computational and Nonlinear Dynamics}, 12(4):044501,
  2017.

\bibitem{santhosh2016discontinuity}
B~Santhosh, S~Narayanan, and C~Padmanabhan.
\newblock Discontinuity induced bifurcations in nonlinear systems.
\newblock {\em Procedia IUTAM}, 19:219--227, 2016.

\bibitem{awrejcewicz2005stick}
J~Awrejcewicz and D~Sendkowski.
\newblock Stick-slip chaos detection in coupled oscillators with friction.
\newblock {\em International Journal of Solids and Structures},
  42(21-22):5669--5682, 2005.

\bibitem{balcerzak2018spectrum}
M~Balcerzak, A~Dabrowski, A~Stefa{\'n}ski, and J~Wojewoda.
\newblock Spectrum of {L}yapunov exponents in non-smooth systems evaluated
  using orthogonal perturbation vectors.
\newblock In {\em MATEC Web of Conferences}, volume 148, page 10003. EDP
  Sciences, 2018.

\bibitem{pikunov2019numerical}
D~Pikunov and A~Stefanski.
\newblock Numerical analysis of the friction-induced oscillator of {D}uffing's
  type with modified {L}u{G}re friction model.
\newblock {\em Journal of Sound and Vibration}, 440:23--33, 2019.

\bibitem{benettin1976kolmogorov}
G~Benetti, L~Galgani, and J~Strelcyn.
\newblock Kolmogorov entropy and numerical experiments.
\newblock {\em Physical Review A}, 14(6):2338, 1976.

\bibitem{grassberger1983characterization}
P~Grassberger and I~Procaccia.
\newblock Characterization of strange attractors.
\newblock {\em Physical Review Letters}, 50(5):346, 1983.

\bibitem{lima2015stick}
R~Lima and R~Sampaio.
\newblock Stick-mode duration of a dry-friction oscillator with an uncertain
  model.
\newblock {\em Journal of Sound and Vibration}, 353:259--271, 2015.

\bibitem{kocarev2006discrete}
L~Kocarev, J~Szczepanski, J~Amig{\'o}, and I~Tomovski.
\newblock Discrete chaos-i: Theory.
\newblock {\em IEEE Transactions on Circuits and Systems I: Regular Papers},
  53(6):1300--1309, 2006.

\bibitem{awrejcewicz2018quantifying}
J~Awrejcewicz, A~Krysko, N~Erofeev, V~Dobriyan, M~Barulina, and V~Krysko.
\newblock Quantifying chaos by various computational methods. part 1: Simple
  systems.
\newblock {\em Entropy}, 20(3):175, 2018.

\bibitem{kleijnen2017regression}
J~Kleijnen.
\newblock Regression and kriging metamodels with their experimental designs in
  simulation: a review.
\newblock {\em European Journal of Operational Research}, 256(1):1--16, 2017.

\bibitem{kleijnen2009kriging}
J~Kleijnen.
\newblock Kriging metamodeling in simulation: A review.
\newblock {\em European journal of operational research}, 192(3):707--716,
  2009.

\bibitem{williams1996gaussian}
C~Williams and C~Rasmussen.
\newblock Gaussian processes for regression.
\newblock In {\em Advances in neural information processing systems}, pages
  514--520, 1996.

\bibitem{clarke2005analysis}
S~Clarke, J~Griebsch, and T~Simpson.
\newblock Analysis of support vector regression for approximation of complex
  engineering analyses.
\newblock {\em Journal of mechanical design}, 127(6):1077--1087, 2005.

\bibitem{park1991universal}
J~Park and I~Sandberg.
\newblock Universal approximation using radial-basis-function networks.
\newblock {\em Neural computation}, 3(2):246--257, 1991.

\bibitem{jiang2017adaptive}
P~Jiang, Y~Zhang, Q~Zhou, X~Shao, J~Hu, and L~Shu.
\newblock An adaptive sampling strategy for kriging metamodel based on
  {D}elaunay triangulation and topsis.
\newblock {\em Applied Intelligence}, pages 1--13, 2017.

\bibitem{krige1951statistical}
D~Krige.
\newblock A statistical approach to some basic mine valuation problems on the
  witwatersrand.
\newblock {\em Journal of the Southern African Institute of Mining and
  Metallurgy}, 52(6):119--139, 1951.

\bibitem{sacks1989design}
J~Sacks, W~Welch, T~Mitchell, and H~Wynn.
\newblock Design and analysis of computer experiments.
\newblock {\em Statistical science}, pages 409--423, 1989.

\bibitem{van2003kriging}
W~Van~Beers and J~Kleijnen.
\newblock Kriging for interpolation in random simulation.
\newblock {\em Journal of the Operational Research Society}, 54(3):255--262,
  2003.

\bibitem{pennestri2016review}
E~Pennestr{\`\i}, V~Rossi, P~Salvini, and P~Valentini.
\newblock Review and comparison of dry friction force models.
\newblock {\em Nonlinear dynamics}, 83(4):1785--1801, 2016.

\bibitem{olsson1998friction}
H~Olsson, K~{\AA}str{\"o}m, C~De~Wit, M~G{\"a}fvert, and P~Lischinsky.
\newblock Friction models and friction compensation.
\newblock {\em Eur. J. Control}, 4(3):176--195, 1998.

\bibitem{marques2016survey}
F~Marques, P~Flores, JC~Claro, and H~Lankarani.
\newblock A survey and comparison of several friction force models for dynamic
  analysis of multibody mechanical systems.
\newblock {\em Nonlinear Dynamics}, 86(3):1407--1443, 2016.

\bibitem{dupont2002single}
P~Dupont, V~Hayward, B~Armstrong, and F~Altpeter.
\newblock Single state elastoplastic friction models.
\newblock {\em IEEE Transactions on automatic control}, 47(5):787--792, 2002.

\bibitem{prandtl1924spannungsverteilung}
L~Prandtl.
\newblock Spannungsverteilung in plastischen {K}{\"o}rpern.
\newblock In {\em Proceedings of the 1st International Congress on Applied
  Mechanics}, pages 43--54, 1924.

\bibitem{dupont2000elasto}
P~Dupont, B~Armstrong, and V~Hayward.
\newblock Elasto-plastic friction model: contact compliance and stiction.
\newblock In {\em American Control Conference, 2000. Proceedings of the 2000},
  volume~2, pages 1072--1077. IEEE, 2000.

\bibitem{de1995new}
C~De~Wit, H~Olsson, K~Astrom, and P~Lischinsky.
\newblock A new model for control of systems with friction.
\newblock {\em IEEE Transactions on automatic control}, 40(3):419--425, 1995.

\bibitem{townsend1987effect}
W~Townsend and Jr~Salisbury.
\newblock The effect of {C}oulomb friction and stiction on force control.
\newblock In {\em Proceedings. 1987 IEEE International Conference on Robotics
  and Automation}, volume~4, pages 883--889. IEEE, 1987.

\bibitem{stribeck1902wesentlichen}
R~Stribeck.
\newblock Die wesentlichen {E}igenschaften der {G}leit-und {R}ollenlager.
\newblock {\em Zeitschrift des Vereines Deutscher Ingenieure}, 46:1341--1348,
  1902.

\bibitem{oseledec1968multiplicative}
V~Oseledec.
\newblock A multiplicative ergodic theorem. {L}iapunov characteristic number
  for dynamical systems.
\newblock {\em Trans. Moscow Math. Soc.}, 19:197--231, 1968.

\bibitem{rosenstein1993practical}
M~Rosenstein, J~Collins, and C~De~Luca.
\newblock A practical method for calculating largest {L}yapunov exponents from
  small data sets.
\newblock {\em Physica D: Nonlinear Phenomena}, 65(1-2):117--134, 1993.

\bibitem{shimada1979numerical}
I~Shimada and T~Nagashima.
\newblock A numerical approach to ergodic problem of dissipative dynamical
  systems.
\newblock {\em Progress of theoretical physics}, 61(6):1605--1616, 1979.

\bibitem{benettin1980lyapunov}
G~Benettin, L~Galgani, A~Giorgilli, and J~Strelcyn.
\newblock Lyapunov characteristic exponents for smooth dynamical systems and
  for {H}amiltonian systems; a method for computing all of them. part 1:
  Theory.
\newblock {\em Meccanica}, 15(1):9--20, 1980.

\bibitem{kantz1994robust}
H~Kantz.
\newblock A robust method to estimate the maximal {L}yapunov exponent of a time
  series.
\newblock {\em Physics letters A}, 185(1):77--87, 1994.

\bibitem{wolf1986quantifying}
A~Wolf.
\newblock Quantifying chaos with {L}yapunov exponents.
\newblock {\em Chaos}, 16:285--317, 1986.

\bibitem{molaie2013simple}
M~Molaie, S~Jafari, J~Sprott, and S~Golpayegani.
\newblock Simple chaotic flows with one stable equilibrium.
\newblock {\em International Journal of Bifurcation and Chaos}, 23(11):1350188,
  2013.

\bibitem{matheron1963principles}
G~Matheron.
\newblock Principles of geostatistics.
\newblock {\em Economic geology}, 58(8):1246--1266, 1963.

\bibitem{stein1991universal}
A~Stein and LCA Corsten.
\newblock Universal kriging and cokriging as a regression procedure.
\newblock {\em Biometrics}, pages 575--587, 1991.

\bibitem{handcock1993bayesian}
M~Handcock and M~Stein.
\newblock A {B}ayesian analysis of kriging.
\newblock {\em Technometrics}, 35(4):403--410, 1993.

\bibitem{matern1960spatial}
B~Mat{\'e}rn.
\newblock Spatial variation: Meddelanden fran statens skogsforskningsinstitut.
\newblock {\em Lecture Notes in Statistics}, 36:21, 1960.

\bibitem{dubourg2013metamodel}
V~Dubourg, B~Sudret, and F~Deheeger.
\newblock Metamodel-based importance sampling for structural reliability
  analysis.
\newblock {\em Probabilistic Engineering Mechanics}, 33:47--57, 2013.

\bibitem{bouhlel2019gradient}
M~Bouhlel and J~Martins.
\newblock Gradient-enhanced kriging for high-dimensional problems.
\newblock {\em Engineering with Computers}, 35(1):157--173, 2019.

\bibitem{toal2011development}
D~Toal, N~Bressloff, A~Keane, and C~Holden.
\newblock The development of a hybridized particle swarm for kriging
  hyperparameter tuning.
\newblock {\em Engineering optimization}, 43(6):675--699, 2011.

\bibitem{santner2013design}
T~Santner, B~Williams, and W~Notz.
\newblock {\em The design and analysis of computer experiments}.
\newblock Springer Science \& Business Media, 2013.

\bibitem{jones1998efficient}
D~Jones, M~Schonlau, and W~Welch.
\newblock Efficient global optimization of expensive black-box functions.
\newblock {\em Journal of Global optimization}, 13(4):455--492, 1998.

\bibitem{turner2007multidimensional}
Cameron~J Turner, Richard~H Crawford, and Matthew~I Campbell.
\newblock Multidimensional sequential sampling for {NURB}s-based metamodel
  development.
\newblock {\em Engineering with Computers}, 23(3):155--174, 2007.

\bibitem{singh2013balanced}
P~Singh, D~Deschrijver, and T~Dhaene.
\newblock A balanced sequential design strategy for global surrogate modeling.
\newblock In {\em Simulation Conference (WSC), 2013 Winter}, pages 2172--2179.
  IEEE, 2013.

\bibitem{liu2017adaptive}
H~Liu, J~Cai, and Y~Ong.
\newblock An adaptive sampling approach for kriging metamodeling by maximizing
  expected prediction error.
\newblock {\em Computers \& Chemical Engineering}, 106:171--182, 2017.

\bibitem{sundararajan2000predictive}
S~Sundararajan and S~Keerthi.
\newblock Predictive approaches for choosing hyperparameters in {G}aussian
  processes.
\newblock In {\em Advances in neural information processing systems}, pages
  631--637, 2000.

\bibitem{lam2008sequential}
C~Lam.
\newblock {\em Sequential adaptive designs in computer experiments for response
  surface model fit}.
\newblock PhD thesis, The Ohio State University, 2008.

\bibitem{Class_Krig_sub}
JN~Fuhg and A~Fau.
\newblock An innovative adaptive kriging approach for efficient binary
  classification of mechanical problems.
\newblock {\em manuscript submitted to publication}, in 2019.

\bibitem{Jan_Master_thesis}
JN~Fuhg.
\newblock Adaptive surrogate models for parametric studies.
\newblock Master's thesis, Leibniz Universit\"{a}t Hannover, Arxiv platform
  https://arxiv.org/abs/1905.05345, 2019.

\bibitem{viana2010algorithm}
F~Viana, G~Venter, and V~Balabanov.
\newblock An algorithm for fast optimal latin hypercube design of experiments.
\newblock {\em International journal for numerical methods in engineering},
  82(2):135--156, 2010.

\end{thebibliography}

\end{document}